\documentclass[oldversion]{aa}

\usepackage{graphicx}
\usepackage{natbib}

\bibpunct{(}{)}{;}{a}{}{,}

\usepackage{txfonts}

\begin{document}

   \title{The outflow in Mrk 509}

   \subtitle{A method to calibrate \textit{XMM-Newton} EPIC-pn and RGS}

   \author{R.G. Detmers\inst{1}
          \and
          J.S. Kaastra\inst{1}\inst{,2}
          \and 
          E. Costantini\inst{1}
          \and
          F. Verbunt\inst{2}
          \and
          M. Cappi\inst{3}
           \and
          C. de Vries\inst{1}}

   \offprints{R.G.Detmers}

   \institute{SRON Netherlands Institute for Space Research, Sorbonnelaan 2, 3584 CA Utrecht, The Netherlands \email{r.g.detmers@sron.nl}
    \and
    Astronomical Institute, University of Utrecht, Postbus 80000, 3508 TA Utrecht, The Netherlands
    \and
    INAF-IASF Bologna, Via Gobetti 101, I-40129 Bologna, Italy}     

   \date{}

 
  \abstract{We have analyzed three \textit{XMM-Newton} observations of the Seyfert 1 galaxy Mrk 509, with the goal to detect small variations in the ionized outflow properties. Such measurements are limited by the quality of the cross-calibration between RGS, the best instrument to characterize the spectrum, and EPIC-pn, the best instrument to characterize the variability. For all three observations we are able to improve the relative calibration of RGS and pn consistently to 4 \%. In all observations we detect three different outflow components and, thanks to our accurate cross-calibration we are able to detect small differences in the ionization parameter and column density in the highest ionized component of the outflow. This constrains the location of this component of the outflow to within 0.5 pc of the central source. Our method for modeling the relative effective area is not restricted to just this source and can in principle be extended to other types of sources as well.}

   \maketitle
%

\section{Introduction}			\label{intro}

Active Galactic Nuclei (AGN) outflows are thought to play an important role in feedback processes, which connect the growth of the black hole to the growth of the galaxy \citep{DiMatteo05}. However the strength of these outflows (kinetic luminosity) is unknown without knowing their precise distance from the central source. In the UV density sensitive line diagnostics have been used to constrain the density of the outflowing gas and hence also the distance \citep{Gabel03}. Attempts have been made to do the same in the X-ray regime \citep{Kaastra04}, but these results were affected by large uncertainties. Therefore in the X-ray band currently the best way to determine the distance of the X-ray outflows is by using variability combined with high-resolution X-ray spectroscopy to constrain the density of the gas and then use this constraint to put an upper limit to the distance $R$. The main challenge with this approach is that it requires several crucial ingredients:

\begin{itemize}
\item  High-resolution X-ray spectroscopic observations to determine the ionization state, column densities and outflow velocities of the warm absorber
\item  Multiple observations of the same source
\item  A source that varies on a suitable timescale and changes significantly in ionizing flux or in warm absorber properties
\end{itemize}
If the source varies too fast, then it becomes difficult to obtain time-resolved spectra suitable to ionized outflow studies. If the changes in ionizing flux are small, then detecting changes in high resolution X-ray data is very difficult. There are several sources for which the variability approach has been successful, all with difficulties or with large changes in flux, however. The \textit{Chandra} data of NGC 3783 \citep{Netzer03, Krongold05} were spread out over several months (with distance limits between 0.2 and 25 pc), while the \textit{XMM-Newton} observations of this source \citep{Behar03} occurred during a gradual rising phase without clear flux extrema, with no changes in the absorber properties leading to a minimum distance of 0.5 - 2 pc from the source, depending on the ionization component. In NGC 4051 the flux dropped by a factor of five \citep{Steenbrugge09} leading to a 0.02 - 1 pc upper limit on the location of the absorbing gas. For NGC 5548 there was a 3 year gap in the observations \citep{Detmers08} and the flux changed by a factor of five in the soft (0.2$-$2 keV) band, which resulted in an upper limit of 7 pc to the location of the warm absorber. Also in NGC 3516 variability was used to constrain the location of the ionized outflow \citep[less than 0.2 pc from the source][]{Netzer02}, but the flux change was very large (50 at 1 keV).

In principle, the \textit{XMM-Newton} EPIC-pn instrument \citep{Struder01} combined with the \textit{XMM-Newton} Reflection Grating Spectrometer \citep[RGS,][]{denHerder01} can be used to detect small changes in the column density of the ionized outflow. The RGS spectrum gives the detailed warm absorber structure such as outflow velocity, velocity width and ionization state. EPIC-pn with its larger effective area can detect small variations in the column density in response to small changes in the continuum flux. This requires that the relative calibration between EPIC-pn and RGS is better known than the observed variations in the column density and ionization parameter.

To this end we have analyzed three different observations of the bright Seyfert 1 galaxy Mrk 509 \citep[z = 0.0344, $N_{\mathrm{H}}$ = 4.4 $\times$ 10$^{24}$ m$^{-2}$,][]{Fisher95,Murphy96}, taken by \textit{XMM-Newton} EPIC-pn and RGS, during different epochs in order to search for any variability in the outflow. We investigate whether we can model the relative calibration between EPIC-pn and RGS in a consistent way, in order to take it into account when fitting the EPIC-pn and RGS spectra simultaneously. If successful, this approach allows for variability studies of weakly varying AGN, enabling us to constrain the outflow location also in these sources. Additionally this approach can be extended to other types of sources as well, as the method itself is source independent even though the exact model of the relative calibration will depend on the source type. We stress that it is also important, to take into account that the calibration of the instruments (e.g. effective area) has changed in time.

Mrk 509 has been observed with high-resolution gratings before, namely in 2000 and 2001 by \textit{XMM-Newton} \citep{Pounds01,Page03} and \textit{Chandra} HETGS \citep{Yaqoob03}. \citet{Smith07} (hereafter S07) re-analyzed the RGS data and found evidence of three warm absorber components, as well as broad and narrow emission lines. We use this study as a starting point of our analysis of the three new observations taken in 2005 and 2006.
Using these data, \citet{Cappi09} found evidence of a highly ionized and mildly relativistic outflow. Clearly Mrk 509 is an excellent source to investigate in further detail.

The most recent calibrations of the RGS and pn instruments \citep[see][for the RGS]{Kaastra09} show that the uncertainty in the effective area of the RGS can be reduced to 3 $\%$, while the pn absolute effective area calibration has an accuracy of 10 $\%$\footnote{http://xmm2.esac.esa.int/docs/documents/CAL-TN-0018.pdf}. Our aim is to obtain a more accurate relative calibration for RGS and pn.

\begin{table*}
\caption{Mrk 509 Observations.}             
\label{tab:obs}      
\centering                          
\begin{tabular}{l c c c c c c}        
\hline\hline                 
Observation & Date & Obsid & Duration & Exposure RGS$^{a}$ & Exposure pn$^{a}$ & Flux$^{b}$  \\    
                       &	        &	     & (ks) & (ks) & (ks) & (W m$^{-2}$ s$^{-1}$) \\
\hline                        
   1 & 18-10-2005 & 0306090201 & 86 & 85 & 58 & 7.1 $\times$ 10$^{-14}$  \\      
   2 & 20-10-2005 & 0306090301 & 47 & 47 & 32 & 7.2 $\times$ 10$^{-14}$  \\
   3 & 24-04-2006 & 0306090401 & 70 & 67 & 43 & 8.1 $\times$ 10$^{-14}$  \\
\hline                                   
\multicolumn{7}{l}{$^a$ Effective exposure time after filtering.} \\
\multicolumn{7}{l}{$^b$ The flux is given for the pn 0.2$-$10 keV band.} \\
\end{tabular}
\end{table*}

Sect. \ref{data} describes the data reduction we have applied to all three observations. In Sect. \ref{method} we describe the method we used to determine the cross-calibration differences between pn and RGS and how we model these. In Sect. \ref{cross} we present the results we have obtained for the cross-calibration. Sect. \ref{spectral} contains the spectral analysis of all three spectra. We discuss our results in Sect. \ref{discussion} and present our conclusions in Sect. \ref{conclusions}.  

\section{Data reduction}				\label{data}

The log of the three observations of Mrk 509 that we used is given in Table \ref{tab:obs}. We have used the SAS 9.0 software package to reduce all the data. We only use these three observations, as the S-N ratio of the first two RGS observations (SN of 3 and 4 in the 7$-$38 $\AA$ respectively) is too low for this study due to a high background level. The S-N ratios of the three observations used here are 9, 7 and 8 respectively. We do not use the MOS data here, which are affected by pile-up and therefore their useful part has poorer S$-$N ratio as compared to the pn data.

For the RGS spectra we use the \textit{rgsproc}\footnote{http://xmm.vilspa.esa.es/sas/8.0.0/doc/rgsproc/index.html} task with the source coordinates as obtained from SIMBAD\footnote{http://simbad.u-strasbg.fr/simbad} ($\alpha$ = 311.04067 degrees, $\delta$ = -10.72353 degrees). After checking the background lightcurve for possible flaring intervals (observation 3 had one event with background level 0.25 cts/s and observations 5 had two events with background 0.4 cts/s, the mean net source count rate being on average 2.8 cts/s), we determine the good time intervals and rerun the \textit{rgsproc} task now with the good time intervals selected (using a cutoff of 0.1 cts/s in the background lightcurve) and also with the \textit{keepcool} keyword set to \textit{no}, so cool pixels are also rejected. We also stacked all three observations with the \textit{rgscombine} task for both RGS1 and RGS2 in order to compare the results of the three individual observations with the stacked spectrum. 

The OM data were used to obtain fluxes in the UV band, which were used along with the pn data to construct the spectral energy distribution (SED) of Mrk 509. The fluxes in the UVW2, UVW1 and UVM2 filters were obtained by extracting the magnitudes from the PPS products of the OM data. These were then corrected for the extinction due to the Galactic absorption towards the source, using the relation between $N_{\mathrm{H}}$ and A$_{\mathrm{V}}$ from \citet{Predehl95} and the relation between A$_{\mathrm{\lambda}}$ and A$_{\mathrm{V}}$, as given by \citet{Cardelli89}

All the pn data were taken in small window mode with a thin filter. The background lightcurves were checked for flaring intervals and the good time intervals were determined to produce a filtered dataset. For the pn data we have first checked for possible pile-up effects. Using the \textit{epatplot} task we analyzed the number of single and double events and found no signs of pile-up in our spectra. We also checked for X-ray loading, which can occur even when the countrate is below the pile-up limit. X-ray loading is the inclusion of X-ray events in the pn offset map which is produced at the start of the observation. This leads to additional and incorrect offset shifts for pixels where the X-ray loading occurs. All events in the subsequent observation associated with that pixel will be shifted to lower energies, depending on the energy of the incoming X-ray photons. A more detailed description can be found in the EPIC calibration online documentation, document TN-0050\footnote{http://xmm2.esac.esa.int/docs/documents/CAL-TN-0050-1-0.ps.gz}. 
In order to counteract the effects of X-ray loading we excluded the central region from our analysis. We therefore took an annular extraction region centered around the source with an inner radius of 50 (2.5 arcsec) and an outer radius of 600 (30 arcsec) in detector coordinates. The background was extracted from an annulus of the same size outside the source region and avoiding out of time events. We use the SPEX\footnote{http://www.sron.nl/spex} spectral fitting package \citep{Kaastra96} and we use C-statistics for fitting the spectrum. The errors are calculated for $\Delta$\,C = 1.    

\section{The relative calibration between pn and RGS}			\label{method}

\subsection{Setup}			\label{setup}

We only use data between 0.32 keV and 1.77 keV (this corresponds to the RGS band, i.e. 7$-$38 $\AA$). We follow a method in four steps, detailed below, to determine the cross-calibration between pn and RGS. We first fit a model spectrum to the RGS data. All model parameters are frozen and the model is then applied to the pn data. The ratios of predicted to observed pn count rates are then modeled to determine the relative calibration between RGS and pn. The last step is to do a simultaneous fit of both the pn and RGS spectrum, taking the relative calibration, i.e.  into account.
As an additional check we created a fluxed RGS spectrum of each observation and follow the same procedure as above. 

\subsection{Spectral Models}				\label{spectral_models}

We model all RGS spectra with a power-law and blackbody component, absorbed by three photoionized components that are modelled using the \textit{xabs} model of SPEX. Also the cosmological redshift and Galactic absorption are taken into account. Broad emission lines are modeled with a Gaussian line profile, narrow emission lines with a delta line profile. The flux at earth $f_\lambda$ may then be written as:

\begin{eqnarray}							\label{eq:model}
f_\lambda &=& \left(R(z) \otimes \left( f_\mathrm{PL}(F_\mathrm{PL},\Gamma_\mathrm{PL})
          +  f_\mathrm{MBB}(F_\mathrm{MBB},T_\mathrm{MBB})
 +  \sum_{i=1}^2 G_i(F_i,\lambda_i,\sigma_i)\right. \right. \nonumber \\
 &+&\left. \left. \sum_{\Delta=1}^1 f_\Delta(\lambda_\Delta,F_\Delta)\right)
\times \prod_{j=1}^3 X_i(N_j,\xi_j,v_j,\sigma_j) \right ) A(N_\mathrm{H})
\end{eqnarray}

We normalize the fluxes of the various components by their total flux in the RGS band. 
The power-law $f_{\mathrm{PL}}$ is described by a normalization $F_{\mathrm{PL}}$ and a slope $\Gamma_{\mathrm{PL}}$, while the MBB $f_{\mathrm{MBB}}$ is described by a normalization $F_{\mathrm{MBB}}$ and an effective temperature $T_{\mathrm{MBB}}$ (see Table \ref{tab:cont}). 
Each broad emission line $G_{\mathrm{i}}$ is characterized by a normalization $F_{i}$, central wavelength $\lambda_{i}$, and width $\sigma_{i}$, while the narrow line $f_{\mathrm{\Delta}}$ is characterized by a normalization $F_{\Delta}$ and wavelength $\lambda_{\Delta}$.
Each absorption component $X_{}$ is characterized by a hydrogen column density $N_{j}$, ionization parameter $\xi_{j}$, outflow velocity $v_{j}$ and Gaussian velocity broadening $\sigma_{j}$.
The effect of cosmological redshift $R$ depends on redshift $z$. The galactic absorption $A$ depends on the hydrogen column density $N_{\mathrm{H}}$.

The widths $\sigma_{i}$ of the two broad emission lines are kept fixed to the values determined from the optical observations and the central wavelengths $\lambda_{i}$ are fixed to the laboratory wavelengths (see Table \ref{tab:lines}). The high wavelength resolution of the RGS requires a detailed description of the wavelength dependence of the interstellar absorption. For this reason we use the SPEX model \textit{hot} with a fixed low temperature $kT$ = 5 $\times$ 10$^{-4}$ keV and a fixed column density determined in earlier observations $N_{\mathrm{H}}$ = 4.4 $\times$ 10$^{24}$ m$^{-2}$ (Sect \ref{intro}).

\begin{figure}[htbp]
   \includegraphics[angle= -90,width=8cm]{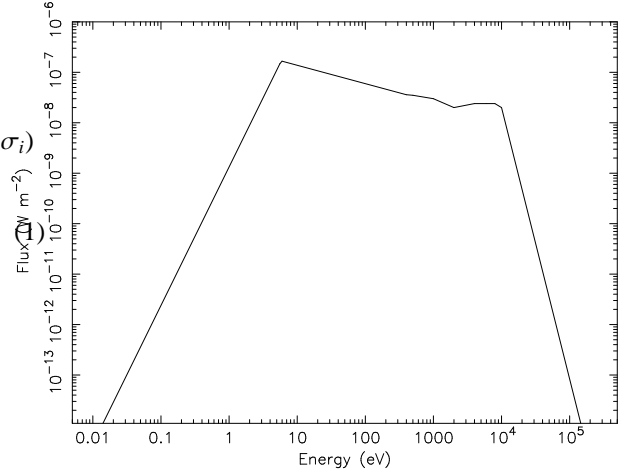}
   \caption{\label{fig:sed}
        The SED used for the Mrk 509 observations. }
\end{figure}

The ionization states $\xi_{j}$ of the photoionized absorbers depend strongly on the assumed SED involved \citep{Chakravorty09}. We have constructed the SED from the Mrk 509 observations, using the pn and OM data to obtain the necessary flux points. The SED's of the three observations are almost identical, so we use the SED obtained from the first observation. The photoionization balance calculations needed for our spectral modeling were based on this SED and performed using version C08.00 of Cloudy\footnote{http://www.nublado.org/}, an earlier version is described by \citet{Ferland98}. The SED used is shown in Fig. \ref{fig:sed}.

\subsection{Method 1}

The first step is to obtain an accurate RGS fit for each observation. This is done by fitting the spectral model described in the previous subsection (see Eq. \ref{eq:model}) to the RGS data.
The second step is that we use this fit with all parameters fixed to predict the count rate $C_{\mathrm{M}}(PI)$ in the pn. 
We then compare these with the observed $C_{\mathrm{O}}(PI)$ count rates in the pn by computing:
\begin{equation}		\label{eq:ratio}
\centering
      R(PI) = \frac{C_{\mathrm{O}}(PI)}{C_{\mathrm{M}}(PI)}	
\end{equation}
The resulting values for $R(PI)$ are plotted in Fig. \ref{fig:resepic}. 

As the third step we calculate these ratios $R(PI)$ using the SPEX model \textit{knak} $F$. This is a piece-wise broken power-law, which consists of hinge points ($\lambda_{1}$, $\lambda_{2}$, .... ,$\lambda_{\mathrm{k}}$) and the corresponding transmission values ($t_{1}$, $t_{2}$, ....., $t_{\mathrm{k}}$). We chose the hinge points by eye for the first observation, and keep these fixed for all other observations. The transmission values $t_{i}$ are then fitted to the ratio $R$ for each observation, leading to the values shown in Table \ref{tab:knak}. 

At this point we check the \textit{knak} model $F$ as follows: we compute $F\,\times\,f_\lambda$ and fold the result through the response matrix of the pn to predict the model count rate, and compare this with the observed pn count rate. Systematic deviations would indicate that the \textit{knak} model must be revised, e.g. by redefining the $\lambda_{k}$. We find that the \textit{knak} model given in Table \ref{tab:knak} does not produce such systematic deviations (see Fig. \ref{fig:knakres}). 

Now that we have a model $F$ for the cross-calibration, we can fit the simultaneous RGS + pn spectrum. This is the fourth and final step. We create two sectors in SPEX, so that we can have two sets of models for the pn and RGS datasets, namely $f_{\mathrm{\lambda}}$ for the RGS and $f_{\mathrm{pn}} \equiv F f_\lambda$ for the pn. 
We then couple the model parameters in such a way to take advantage of the strengths of each instrument. We let the RGS spectrum determine the high-resolution features, such as the fluxes $F_i$ and $F_\Delta$ of the emission lines and the outflow velocity $v_{j}$ and velocity broadening $\sigma_j$ of the ionized absorbers. This is because pn does not have the resolution necessary to accurately determine the detailed structure (broadening + outflow velocity) of the outflow.
The pn spectrum determines the parameters of the PL and MBB, and the column densities $N_{j}$ and ionization parameters $\xi_{j}$ of the ionized absorbers (\textit{xabs}).
The resulting values are shown in Tables \ref{tab:cont}, \ref{tab:lines} and \ref{tab:warm} together with the values of the fixed parameters. 

We use this four step procedure for every observation, including the stacked one. 

\subsection{Method 2}

As an independent check for each observation, we use the best-fit model $f_{\lambda}$ for each observation to predict the RGS count rate $C_{\mathrm{RGS}}$ and compute the model flux to be used for the pn $f_{pn}$ by multiplying $f_{\lambda}$ with the ratio of the observed to predicted RGS count rates $ \frac{C}{C_{\mathrm{RGS}}}\times\,f_{\lambda}$.
We then predict the pn count rate by folding $f_{pn}$ through the corresponding pn response matrix and compare the result with the observed count rate. This procedure makes sure that any (time-dependent) astrophysical modeling problems are taken care of and the end result will be the pure cross-calibration residuals.  We then compare the cross-calibration residuals we obtain with this method with the first method and whether the results are consistent with each other (see Fig. \ref{fig:resepic}).

\subsection{Further checks}				\label{cross}

\begin{figure}[tbp]
   \includegraphics[angle= -90,width=8cm]{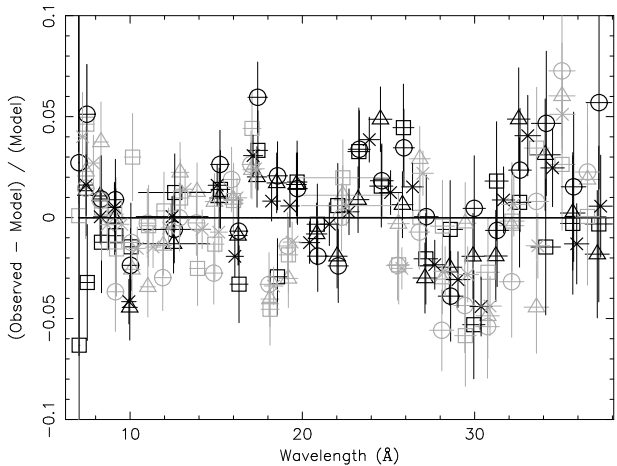}
   \caption{\label{fig:resrgs}
        Broad band RGS fit residuals for all observations. RGS1 data points are indicated by the black symbols, while RGS2 data points are indicated by the gray ones. The different observations are marked by seperate symbols: triangle for observation 1, square for observation 2, circle for observation 3 and cross for the stacked spectrum.}
\end{figure}

To investigate the relative calibration further, we check for the presence of any weak deviations that might bias our RGS fit but would escape attention at full spectral resolution. Fig. \ref{fig:resrgs} shows the residuals in the RGS band after rebinning the spectrum with a factor of 100. The typical residual fluctuations are at the level of 3$-$4$\%$. The third observation shows a discrepancy between the two RGS spectra. The RGS1 data points are consistently below the RGS2 points, with a difference of about 2 \%. This will have an effect on the total RGS flux, which we discuss later in more detail. 

Focusing on the pn spectrum, we load the best fit RGS model into the corresponding pn data set. If the relative calibration would be perfect, then the resulting fit quality of the pn spectrum should be the same as for the RGS only fit. Instead we find something different. The resulting residuals are shown for each observation in Fig. \ref{fig:resepic}. RGS and pn are in agreement with each other within 1$-$4$\%$ below 20 $\AA$. Above 20 $\AA$ the RGS flux is on average 8$-$10$\%$ higher than the pn flux. All three observations and the stacked spectrum are in agreement with each other (apart from small fluctuations on the level of 2 \%).

As mentioned earlier we model these calibration offsets by a \textit{knak} component. The best-fit parameters of this component are shown in Table \ref{tab:knak}. When applying this correction to each pn observation the fit improves dramatically. The resulting residuals are shown in Fig. \ref{fig:knakres}. Some fluctuations on the order of 2 $\%$ can still be seen in the individual spectra, but our model adequately corrects for the large scale offsets. 
We chose the hinge points by eye and fix them for all other observations. The locations of these hinge points were chosen such, that they follow the trend visible in the pn residuals (see Fig. \ref{fig:resepic} in the Appendix). If we instead use a different number of points (e.g. 7 points, with fixed 5 $\AA$ spacing from 7 to 37 $\AA$), the residuals in Fig. \ref{fig:knakres} change by no more than 1$-$2 \% We also performed additional tests with only 5 equidistant hinge points and the results are similar.  

Thus having obtained a model for the relative calibration for each observation, we perform a simultaneous fit of both the pn and RGS spectra, with the relative calibration differences taken into account by our \textit{knak} model.

\begin{table}
\caption{\textit{Knak} model parameters for the cross-calibration residuals. A value lower than 1 means the RGS flux is higher than the pn flux.}             
\label{tab:knak}      
\centering                          
\begin{tabular}{c c c c c}        
\hline\hline                 
$\lambda$ ($\AA$) & 1 & 2 & 3 & stacked \\    
\hline                        
   7   & 0.97$\pm$0.01& 1.00$\pm$0.01& 0.98$\pm$0.01& 0.99$\pm$0.01 \\    
   13 & 0.98$\pm$0.01& 0.97$\pm$0.01& 1.01$\pm$0.01& 0.98$\pm$0.01\\    
   15 & 0.94$\pm$0.01& 0.97$\pm$0.01& 0.97$\pm$0.01& 0.96$\pm$0.01\\    
   19 & 0.96$\pm$0.01& 0.98$\pm$0.01& 0.98$\pm$0.01& 0.97$\pm$0.01\\    
   23 & 0.87$\pm$0.01& 0.92$\pm$0.01& 0.91$\pm$0.01& 0.90$\pm$0.01\\    
   28 & 0.92$\pm$0.01& 0.94$\pm$0.01& 0.91$\pm$0.01& 0.92$\pm$0.01\\    
   34 & 0.79$\pm$0.02& 0.78$\pm$0.02& 0.81$\pm$0.02& 0.80$\pm$0.01\\
   37 & 0.90$\pm$0.03& 0.92$\pm$0.03& 0.94$\pm$0.03& 0.91$\pm$0.02\\    
\hline
\end{tabular}
\end{table}

\section{Spectral Analysis}	\label{spectral}

\subsection{Continuum}		

We have modeled the continuum for all spectral fits with a power-law and a modified blackbody component. For all three observations and the stacked spectrum the continuum model parameters are shown in Table \ref{tab:cont}. If we instead only use a powerlaw to model the continuum as was done in S07, the fits for all three observations are worse by $\Delta$\,C = 75 per d.o.f. Using only the powerlaw as a continuum has a slight effect on the absorption components, as the powerlaw slope is softer in this case ($\Gamma$ = 2.6 instead of 2.3). However given the large change in $\Delta$\,C, we use the continuum model of the powerlaw and modified blackbody from here on.
The spectrum appears to be harder during the 3rd observation, while there is no significant difference between the first two observations (as expected, since they were taken only two days apart). The continuum parameters for the simultaneous fit are exactly the same as the RGS fits, but the errors are reduced due to the addition of the pn spectrum.

\begin{table}
\caption{Continuum parameters for all observations.}             
\label{tab:cont}      
\centering                          
\begin{tabular}{c c c c c}        
RGS \\
\hline\hline                 
Observation & F$_{\mathrm{PL}}$$^{a}$ & $\Gamma_{\mathrm{PL}}$ & F$_{\mathrm{MBB}}$$^{a}$ & kT$_{\mathrm{MBB}}$ (keV) \\    
\hline                        
   1 & 5.42$\pm$0.07 & 2.30$\pm$0.03 & 0.59$\pm$0.11 & 0.092$\pm$0.006 \\      
   2 & 5.73$\pm$0.08 & 2.35$\pm$0.03 & 0.67$\pm$0.17 & 0.093$\pm$0.007 \\
   3 & 4.90$\pm$0.10 & 2.06$\pm$0.04 & 0.91$\pm$0.11 & 0.129$\pm$0.006 \\      
 all & 5.44$\pm$0.05 & 2.26$\pm$0.02 & 0.68$\pm$0.09 & 0.096$\pm$0.004 \\
\hline
RGS + pn \\
\hline\hline                 
Observation & F$_{\mathrm{PL}}$$^{a}$ & $\Gamma_{\mathrm{PL}}$ & F$_{\mathrm{MBB}}$$^{a}$ & kT$_{\mathrm{MBB}}$ (keV) \\    
\hline                        
   1 & 5.41$\pm$0.05 & 2.30$\pm$0.01 & 0.61$\pm$0.06 & 0.092$\pm$0.003 \\      
   2 & 5.72$\pm$0.06 & 2.35$\pm$0.02 & 0.67$\pm$0.11 & 0.093$\pm$0.005 \\
   3 & 4.94$\pm$0.06 & 2.07$\pm$0.02 & 0.87$\pm$0.05 & 0.129$\pm$0.002 \\      
 all & 5.51$\pm$0.02 & 2.27$\pm$0.01 & 0.63$\pm$0.04 & 0.095$\pm$0.002 \\
\hline
\multicolumn{5}{l}{$^a$ The unabsorbed flux in the 0.3$-$2 keV band in 10$^{-14}$ W m$^{-2}$.} \\
\end{tabular}
\end{table}

\subsection{Emission Lines}

Several broad and narrow emission lines are visible in the spectrum (see Fig. \ref{fig:warm}, the \ion{O}{vii} f narrow emission line \citep[rest wavelength 22.101 $\AA$][]{Kelly87}, as well as the broad \ion{O}{vii} recombination line at 21.602 $\AA$ \citep{Engstrom95} and a broad \ion{N}{vii} emission line at 24.78 $\AA$ \citep{Garcia65}. The oxygen lines were already detected by S07 (although they detect a broad intercombination line instead of a recombination line), but the \ion{N}{vii} emission line is a newly detected feature. Table \ref{tab:lines} shows the line parameters for the RGS only fit. The wavelength of the \ion{O}{vii} f line appears to shift between observations. However, this region suffers from bad pixels, which vary from observation to observation, so if one of the bad pixels falls on top of the location of the emission line, the fitted position shifts, depending on the surrounding pixels. Since these lines escape detection in the pn spectrum, we do not fit them in the simultaneous spectrum, but take the fitted values from the RGS only spectra. We have fixed the width of the broad emission lines to 0.8 $\AA$ which corresponds to the width of the optical BLR (FWHM $\sim$ 8000 km s$^{-1}$). We also keep the wavelengths fixed to the rest-frame values.
We do not detect any significant variability in the emission lines. 

\begin{table}
\caption{Emission line parameters for all observations, fluxes are corrected for Galactic absorption.}             
\label{tab:lines}      
\centering                          
\begin{tabular}{c@{\,}c@{\,}c@{\,}c@{\,}c}        
\hline\hline                 
Line & Observation& Wavelength  & Flux 					& FWHM  \\ 
         &		       &  ($\AA$)        & (ph m$^{-2}$ s$^{-1}$)      &  ($\AA$) \\
\hline                        
   \ion{O}{vii} r & 1 & 21.602 (f)  & 3.0 $\pm$ 0.4 &  0.8 (f)\\      
   \ion{O}{vii} r & 2 & 21.602 (f) & 2.3 $\pm$ 0.6&  0.8 (f)\\      
   \ion{O}{vii} r & 3 & 21.602 (f)  & 2.3 $\pm$ 0.5 &  0.8 (f)\\      
   \ion{O}{vii} r & all & 21.602 (f)  & 2.4 $\pm$ 0.3&  0.8 (f)\\      
\hline                               
   \ion{N}{vii} Ly$\alpha$  & 1 & 24.78 (f) & 1.0 $\pm$ 0.3 & 0.8 (f) \\
   \ion{N}{vii} Ly$\alpha$  & 2 & 24.78 (f) & 0.8 $\pm$ 0.4 & 0.8 (f)\\
   \ion{N}{vii} Ly$\alpha$  & 3 & 24.78 (f) & 1.0 $\pm$ 0.4 & 0.8 (f)\\
   \ion{N}{vii} Ly$\alpha$  & all & 24.78 (f) & 1.1 $\pm$ 0.2 & 0.8 (f)\\
\hline
   \ion{O}{vii} f & 1 & 22.07 $\pm$ 0.01 & 0.6 $\pm$ 0.2 & - \\
   \ion{O}{vii} f & 2 & 22.07 $\pm$ 0.03 & 0.2 $\pm$ 0.2 & - \\
   \ion{O}{vii} f & 3 & 21.99 $\pm$ 0.02 & 1.0 $\pm$ 0.5 & - \\
   \ion{O}{vii} f & all & 21.99 $\pm$ 0.02 & 0.30 $\pm$ 0.11 & - \\   
\hline
\multicolumn{5}{l}{$^1$ The (f) indicates that the parameter was kept fixed.}\\
\end{tabular}
\end{table}

\subsection{Warm absorber}

\begin{table}
\caption{Warm Absorber parameters.}             
\label{tab:warm}      
\centering                          
\begin{tabular}{c@{\,}c@{\,}c@{\,}c@{\,}c@{\,}c}        
RGS \\
\hline\hline                 
Comp & Obs & $N_{\mathrm{H}}$$^{a}$ & log $\xi$$^{b}$             & $v$$^{c}$      & $\Delta v$$^{d}$ \\    
		   &                        & (10$^{24}$ m$^{-2}$) & (10$^{-9}$ W m) & (km s$^{-1}$) & (km s$^{-1}$)  \\
\hline                        
    &1 & 1.35$\pm$0.31 & 0.60$\pm$0.18 & 95$\pm$40 & $-$ 230$\pm$70\\
 1 &2 & 0.77$\pm$0.27 & 0.34$\pm$0.20 & 95 (f)& $-$ 230 (f)\\
    &3 & 0.61$\pm$0.21 & 0.60$\pm$0.23 & 95 (f)& $-$ 230 (f)\\
    &all & 1.0$\pm$0.2 & 0.6$\pm$0.1 & 170$\pm$ 40& $-$ 120$\pm$ 60\\
\hline                                 
    &1 & 12.0$\pm$2.1 & 1.97$\pm$0.03 & 80$\pm$20& 0$_{-100}^{+20}$\\      
 2 &2 & 11.4$\pm$2.5 & 1.94$\pm$0.03 & 80 (f) & 0 (f)\\      
    &3 & 6.1$\pm$1.7 & 1.91$\pm$0.04 & 80 (f) &  0 (f)\\      
    &all &10.5$\pm$2.0 & 1.95$\pm$0.02 & 70$\pm$ 10& 0$\pm$40\\ 
\hline
    &1 & 52$\pm$21 & 3.27$\pm$0.09 & 65$\pm$50 & $-$ 490$\pm$120\\
 3 &2 & 26$\pm$12 & 3.13$\pm$0.12 & 65 (f) & $-$ 490 (f)\\
    &3 & 23$\pm$9  & 3.06$\pm$0.09 & 65 (f) & $-$ 490 (f)\\ 
    &all & 80$_{-20}^{+90}$ & 3.20$\pm$0.08 & 25$\pm$10 & $-$ 290$\pm$40\\        
\hline
RGS + pn \\
\hline                 
    &1 & 1.36$\pm$0.23 & 0.64$\pm$0.15 & 95$\pm$40 & $-$ 230$\pm$70 \\
 1 &2 & 0.79$\pm$0.22 & 0.35$\pm$0.17 & 95 (f) & $-$ 230 (f)\\
    &3 & 0.68$\pm$0.19 & 0.63$\pm$0.20 & 95 (f) & $-$ 230 (f)\\
    &all & 1.2$\pm$0.2 & 0.61$\pm$0.10 & 170$\pm$40 & $-$ 120$\pm$60\\
\hline                                 
    &1 & 12.0$\pm$1.7 & 2.01$\pm$0.02 & 80$\pm$20 & 0$_{-110}^{+10}$\\      
 2 &2 & 11.4$\pm$1.6 & 1.96$\pm$0.03 & 80 (f) & 0 (f)\\      
    &3 &  6.1$\pm$2.8 & 1.92$\pm$0.04 & 80 (f) & 0 (f)\\      
    &all &10.5$\pm$2.0 & 2.01$\pm$0.03 & 70$\pm$10 &  0$\pm$40\\ 
\hline
    &1 & 51$\pm$14 & 3.29$\pm$0.04 & 65$\pm$50 & $-$ 475$_{-140}^{+60}$ \\
 3 &2 & 25.4$\pm$6.2 & 3.11$\pm$0.06 & 65 (f) & $-$ 475 (f)\\
    &3 & 18.2$\pm$5.0 & 3.01$\pm$0.06 & 65 (f) & $-$ 475 (f)\\ 
    &all & 69$_{-10}^{+80}$ & 3.19$\pm$0.06 & 25$\pm$10 & $-$ 295$\pm$40\\        
\hline
\multicolumn{6}{l}{$^1$ The (f) indicates that the parameter was kept fixed.}\\
\multicolumn{6}{l}{$^a$ Column density.}\\
\multicolumn{6}{l}{$^b$ Ionization parameter.}\\
\multicolumn{6}{l}{$^c$ r.m.s. velocity broadening}\\
\multicolumn{6}{l}{$^d$ Outflow velocity, a negative velocity corresponds to a blueshift.}\\
\end{tabular}
\end{table}

\begin{figure*}[htbp]
\begin{minipage}[b]{1.0\linewidth}
\centering
\includegraphics[width=12cm, angle = -90]{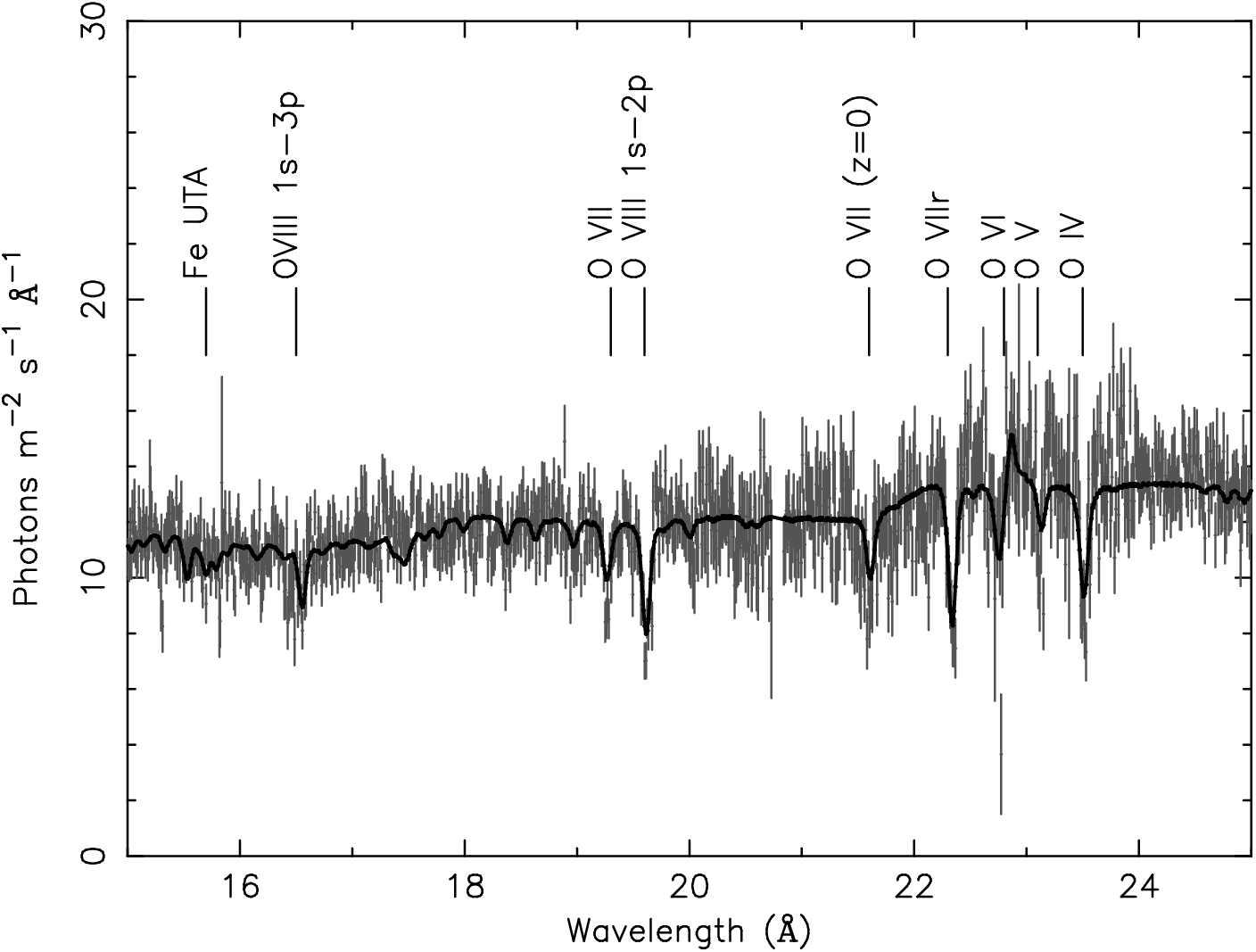}
\end{minipage}
\hspace{2.0cm}
\begin{minipage}[b]{1.00\linewidth}
\centering
\includegraphics[width=12cm, angle = -90]{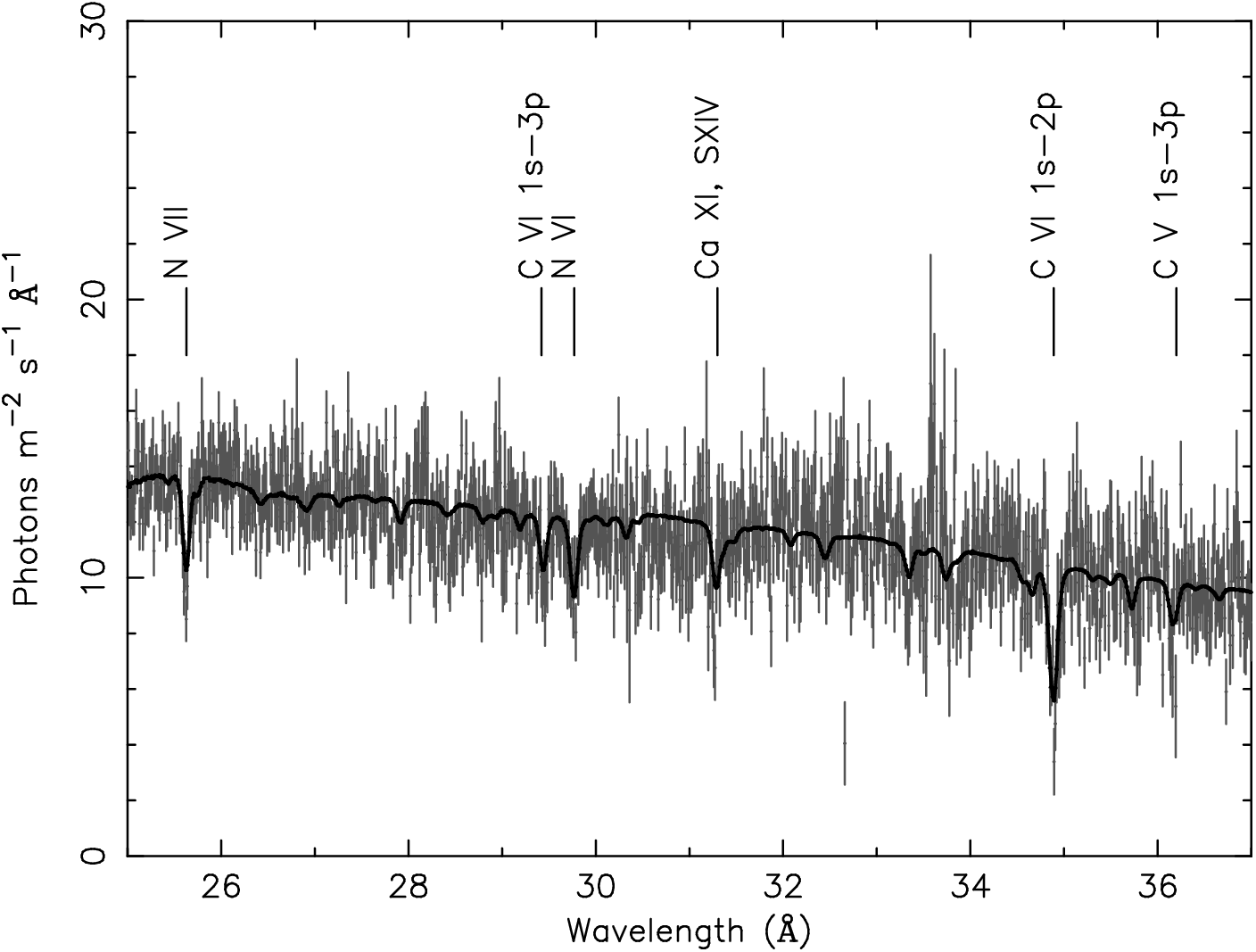}
\caption{\label{fig:warm}
The best fit RGS spectrum for the stacked observation with the absorption lines indicated.}
\end{minipage}
\end{figure*}

We detect three distinct warm absorber components, similar to what S07 have found by stacking two earlier RGS observations of Mrk 509. We started out by using only one component, and then added components as the fit required. The second and third component are needed ($\Delta$\,C = 50 and 40 respectively), but adding a fourth component does not improve the fit any further. The warm absorber in Mrk 509 consists of a low ($\xi$ $\sim$ 0.6), a medium ($\xi$ $\sim$ 2.0) and a high ($\xi$ $\sim$ 3.1) ionization component. The main absorption lines visible in the spectrum can be seen in Fig. \ref{fig:warm}. All warm absorber parameters are shown in Table \ref{tab:warm}. The values for the parameters are consistent for the RGS only and the simultaneous fit, but the errors on the latter are smaller due to the addition of the pn spectra. All parameters were set to a reasonable initial value (based on average values of warm absorber parameters found in the literature) and then freed to obtain the best fit.

\section{Discussion}			\label{discussion}

\subsection{Warm absorber variability}

\begin{figure}[tbp]
   \includegraphics[angle= -90,width=8cm]{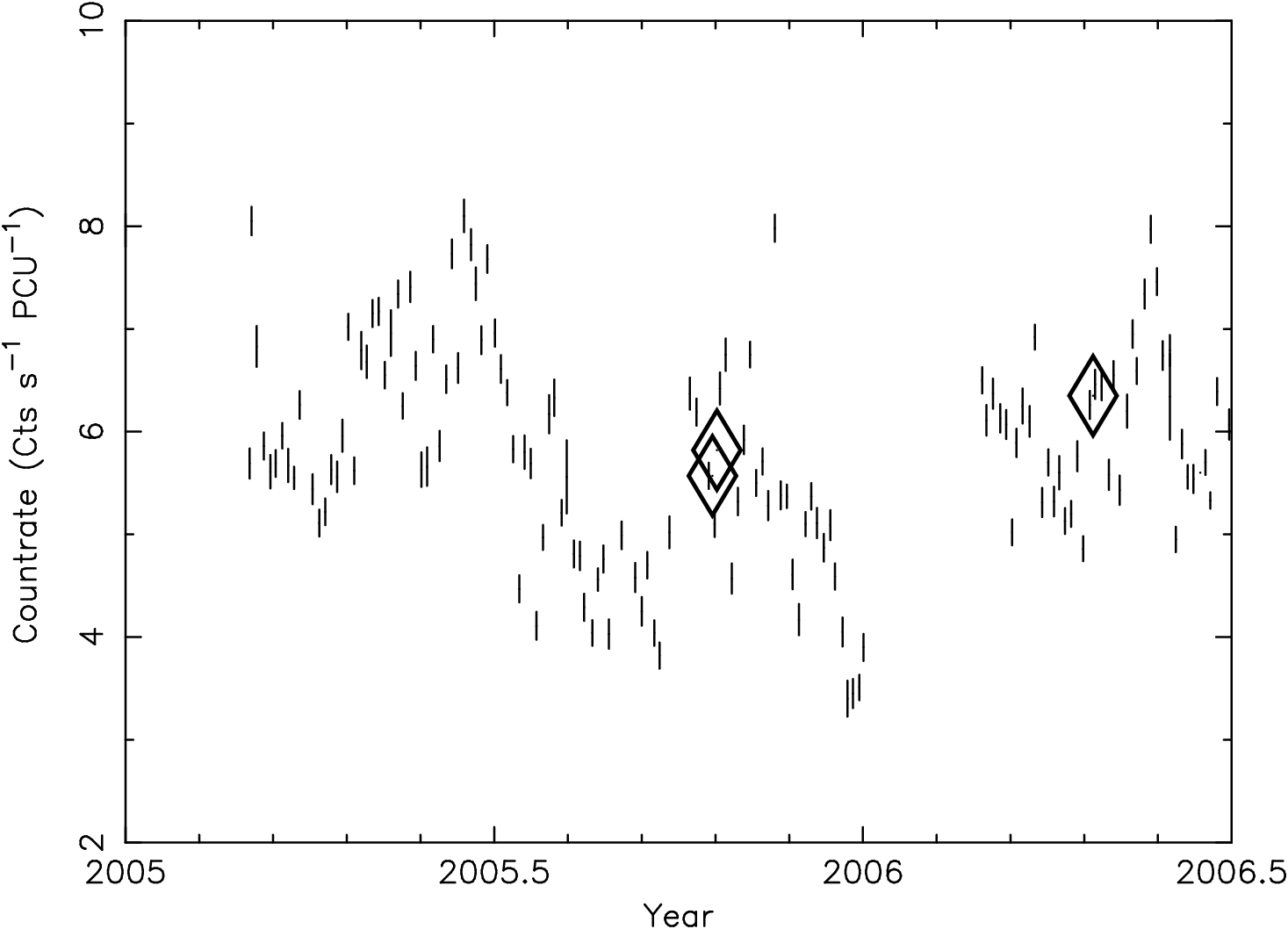}
   \caption{\label{fig:rxte}
        The RXTE 2$-$10 keV lightcurve of Mrk 509, from 2005 to 2007. Data were taken from the online RXTE mission-long data products. The diamonds indicate the exact date of the three \textit{XMM-Newton} observations used in this paper.}
\end{figure} 

\begin{figure}[tbp]
   \includegraphics[angle= -90,width=8cm]{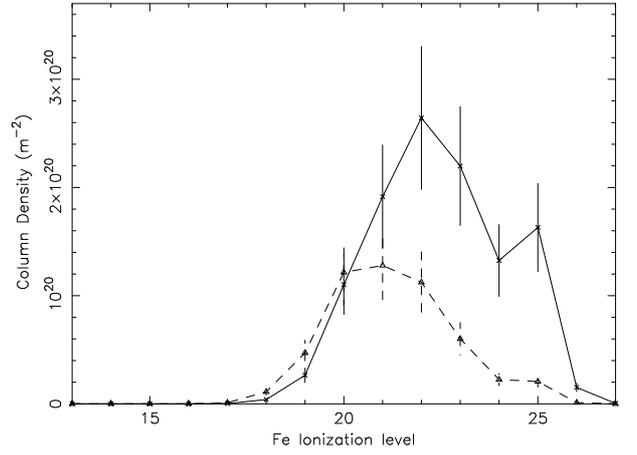}
   \caption{\label{fig:iron}
Column densites for the different ionization levels of iron for the high (log $\xi$ = 3.2) component. The solid line represents the first observation, while the dotted line represents the third observation.}
\end{figure}

Comparing the warm absorber components for all three observations we do not detect any significant variability in components 1 and 2. The high ionization component however shows a change in column density and ionization parameter between the first and the third observation. Changes in column density can be either due to motion across the line of sight or as in NGC 5548 \citep{Detmers08} due to recombination / re-ionization of the absorbing gas due to a change in the continuum flux. 
One has to be careful however in this case, as there is a correlation between $\xi$ and column density, in the sense that a lower $\xi$ and lower column density can produce the same spectral fit as a higher $\xi$, larger column density case. In order to check for this effect we have performed also a fit where we have kept the column density fixed to the column density of the first observation. All three observations have the same $\xi$ value for the high ionization component (log $\xi$ = 3.25). However for the third observations the fit in this case is worsened by $\Delta$\,C = 37 for 1 d.o.f. with respect to the fit shown in Table \ref{tab:warm}. This indicates that the change in ionization parameter and column density between observation 1 and 3 is significant.  
Using the \textit{RXTE} monitoring data of Mrk 509 taken with the PCA, we in principal can know what the continuum flux history is of this source before and after the three $XMM-Newton$ observations. However the gap during the first few months of 2006, prevent us from obtaining a more accurate timescale for the changes detected in the spectrum than the time between the 2nd and 3rd observation, namely half a year.
If we make the assumption that the change in column density and ionization parameter we see are due to recombination of the component 3 gas due to the minimum, then we can get an estimate for the distance of this component using the following relation: 
\begin{equation}		\label{xi}
\centering
      \xi = \frac{L}{n R^{2}}. 
\end{equation}
$\xi$ is the ionization parameter, $L$ the 1 $-$ 1000 Rydberg luminosity, $n$ the density of the gas and $r$ the distance from the ionizing source. The 1 $-$ 1000 Rydberg luminosity of Mrk 509 is 4.9 $\times$ 10$^{37}$ W (as taken from the first observation). If we know the density of the gas then we can constrain the location of the gas, since $\xi$ and $L$ are known from observations. In order to determine $n$, we must know the recombination or ionization timescale of the gas, which is given by \citep{Krolik95,Bottorff00}:
\begin{equation}		\label{trec}
\centering
      \tau_{\mathrm{rec}}(X_{i}) = \left({\alpha_{\mathrm{r}}(X_{i})n \left[\frac{f(X_{i+1})}{f(X_{i})} - \frac{\alpha_{\mathrm{r}}(X_{i-1})}{\alpha_{\mathrm{r}}(X_{i})}\right]}\right)^{-1}, 
\end{equation}
where $\alpha_{\mathrm{r}}(X_{i})$ is the recombination rate from ion $X_{i+1}$ to ion $X_{i}$ and $f(X_{i})$ is the fraction of element $X$ in ionization state $i$. 
From the time between observation 2 and 3 we get an upper limit to $\tau_{\mathrm{rec}}$ of 0.5 years. If we then take for example the \ion{Fe}{XXII} ion (since this is where the ionization level of iron peaks in column density for this component, see Fig \ref{fig:iron}), we get a lower limit to the density $n$ of 1.3 $\times$ 10$^{11}$ m$^{-3}$. 
From this we can constrain the location of this outflow component to within 0.5 pc of the central source. This does not allow us however to distinguish between a scenario where the warm absorber is launched from the accretion disk (i.e. located at BLR distances) or from the dusty torus. Oxygen, neon, carbon and nitrogen are not produced in significant column densities by this high ionization component, so the only ions for which the column densities showed any significant change are \ion{Fe}{XXII} to \ion{Fe}{XXV}. If we perform the same excercise for the other ions, we obtain less strict upper limits (up to 4 pc for \ion{Fe}{XXV}). 

For the other two components which do not vary between the three observations, the non-variability can be due to two reasons:
\begin{itemize}
\item{The density is low, so that the recombination time of the gas is longer than the time between observations 3 + 4 and observation 5, i.e. $\tau_{\mathrm{rec}}$ $\ge$ 0.5 years.}
\item{The density is high enough to track the short time variability of the source, say on a few days timescale, as the fluxes of all observations are within 10 $\%$ of each other.}
\end{itemize}

\subsection{Comparison with previous results}

\citet{Cappi09} have found evidence of a highly ionized and variable outflow component using the same pn data we have used. In the RGS data there is no evidence of such an absorber with a mildly relativistic outflow velocity. As the main goal of this paper focusses on the cross-calibration of RGS and pn, we focussed our attention of the pn analysis only in the 7$-$38 $\AA$ wavelength range. This highly ionized, fast outflowing absorber might represent an extreme part of the outflow described here or it might be disconnected given its high outflow velocity.

If we compare the outflow components in our present spectra with those analyzed by S07, we see that the ionization parameters are consistent with each other, although we find slightly lower values. The column density of the low ionized gas (log $\xi$ $\approx$ 0.6) is a factor of 10 higher in the S07 observations. However the velocity broadening $v$ they find is 0 instead of the 105 km s$^{-1}$ that we find. The main difference however is the outflow velocity. We find the highest outflow velocity ($-$425 km s$^{-1}$) for the highly ionized (log $\xi$ $\sim$ 3.1) component, while S07 find the highest outflow velocity for the component with log $\xi$ $\sim$ 0.9.

These differences may be due to a combination of several effects:
\begin{enumerate}
\item{Systematic uncertainty of 7 m$\AA$ (50$-$200 km s$^{-1}$) in the RGS wavelength scale.}
\item{A different continuum modeling in our paper compared to S07.}
\item{Poorer S$-$N ratio in the S07 data.}
\item{Sometimes an almost similar fit can be obtained by lowering the velocity broadening while increasing the column density, for components that are dominated by saturated lines.}
\item{True time variability of the source.}
\end{enumerate}
All this argues for a monitoring campaign where a high data quality will be obtained and the variations of the source are continuously monitored. Such a campaign is now underway.

\subsection{Cross-calibration between RGS and pn}

\begin{figure}[tbp]
   \includegraphics[angle= -90,width=8cm]{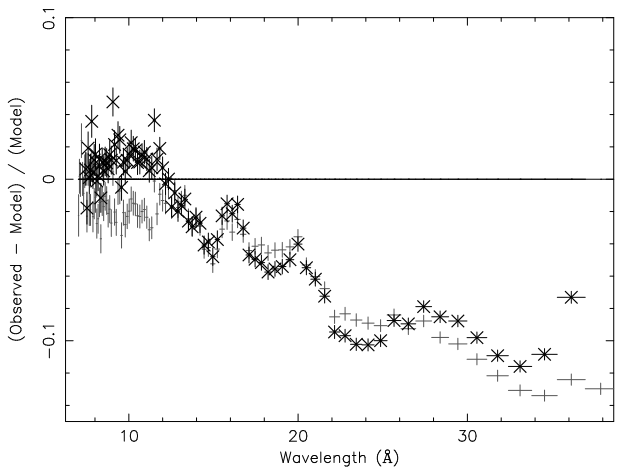}
   \caption{\label{fig:stacked}
      Comparison of the ratios $R$ for both methods (see Fig \ref{fig:resepic}) for the stacked spectrum. The thin points are the $R$ values for method 1 while the thick crosses are the $R$ values for method 2.}
\end{figure}

\begin{figure}[tbp]
   \includegraphics[angle= -90,width=8cm]{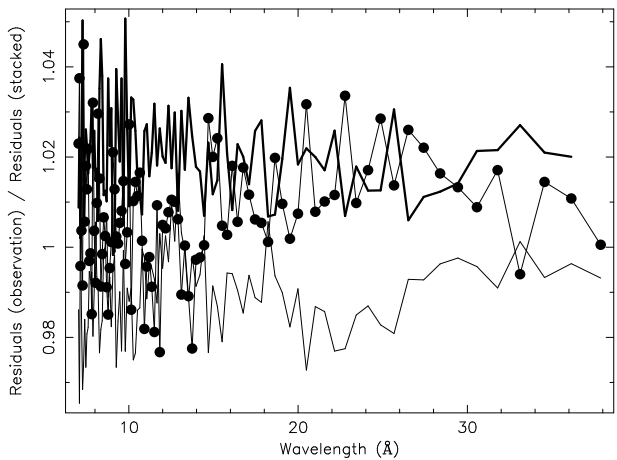}
   \caption{\label{fig:epicres}
      Ratio of the pn residuals for each observation compared to the stacked spectrum. Observation 1 is indicated by the thin black line, observation 2 by the thin black line with dots and observation 3 by the thick black line.}
\end{figure}

This is a first attempt to model the relative calibration between pn and RGS in order to be able to detect small changes in outflow properties. 
Our method for modeling the relative calibration is a relatively simple one, but a cursory glance at the differences between Figs. \ref{fig:resepic} and \ref{fig:knakres} shows the clear improvements for the fit to the pn spectrum.
If we compare the cross-calibration residuals for both the first method (RGS model for pn spectrum) and the second method (fluxed RGS spectrum folded through pn response), we see that apart from a difference for the first observation below 10 $\AA$, the residuals for both methods are almost identical (see Fig. \ref{fig:stacked}). However there is a difference at long wavelengths, particularly from 22 $\AA$ onwards. The same wave-like pattern which can be seen between 22 and 32 $\AA$ in Fig \ref{fig:resrgs}. The difference is about 1.5 \% between the fluxed and the fitted data. The difference between the RGS fluxed and the fitted spectrum can only be due to slight modeling inaccuracies, not due to any calibration offsets in the effective area of RGS or pn, as both methods use the same effective area. This means that by comparing the fluxed and the fitted residuals in the pn, we can separate the inaccuracies due to modeling and those due to uncertainties in the effective area calibration of RGS in the residuals seen in Fig. \ref{fig:resrgs}. These residuals are only apparent when the spectra are heavily rebinned. At higher resolution these residuals would not have been detected. 
From the comparison of the fluxed and fitted pn residuals we find that about 1.5 \% is due to modeling inaccuracies in the RGS fit. The strong wiggle in Fig. \ref{fig:stacked} between 22 and 26 $\AA$ with an amplitude of about 5 \%, may be related to uncertainties in the pn model. Similar features can be seen in the residuals of for instance 3C273 orbit 1381 (see the calibration review tool\footnote{http://xmm2.esac.esa.int/cgi-bin/ept/preview.pl}). 

Both methods show the same general trend for the pn residuals. There is a sharp drop around 23 $\AA$, which is where the oxygen edge is located, possibly indicating that the modeling of this instrumental edge has some uncertainties.
Comparing the residuals in Figs \ref{fig:epicres} and \ref{fig:knakres} between observations, we see that the general shape is exactly the same for each observation. There only appears to be a slight shift of $\sim$ 3 $\%$ between observation 1 and observation 3. The reason for this shift is the difference in flux in the RGS1 and RGS2 spectra of observation 3. Based on the comparison between the \textit{knak} parameters, see Table \ref{tab:knak}, the RGS2 spectrum is the correct one. This means that in Fig. \ref{fig:resepic} the residuals in observation 3 should be shifted by 1$-$2 \% downwards, which brings it into agreement with the previous observations. 
The reason for this 2 \% flux difference is unknown. We have checked for possible differences in exposure times or background filtering criteria between the two RGS, and there is no difference between both RGS exposure times. The difference remains, independent of the exact criteria used for background filtering. As another check we varied the spectral extraction region for the third observation (90, 95 and 98 \% of the cross-disperion PSF respectively). Also here the difference between RGS1 and RGS2 remains. At this moment we have no good explanation for this flux difference. It may however be a rare statistical variation at the 3.5 sigma level. 
 
If we compare the residuals with other cross-calibration investigations, as shown on the calibration review tool website for \textit{XMM-Newton}, we find identical features as in similar sources, like PKS 2155-304 and 3C273 for example, which show that below 0.5 keV the RGS flux is about 10 $\%$ higher than the pn flux. This gives us additional confidence that out method and our results are correct and can be applied to other sources as well. However sources like Mrk 509 are the best to study the cross-calibration, as the pn data are not piled-up, but the source is still bright enough in the RGS to obtain a high-quality spectrum.

The main application for this modeling of the relative calibration is the increased sensitivity to changes in the outflow column densities and the ionization parameter the simultaneous fit provides. 
A comparison of Table \ref{tab:warm} shows that performing a simultaneous fit improves the accuracy of the measurements (column density and ionization state) by almost a factor of two. Without correcting for the relative calibration, this would not have been possible. 

Our method has several assumptions however. The first is that the RGS data give us the best fit for the warm absorber and the emission lines. Also we assume that the RGS continuum fit gives the correct continuum. But since we are interested only in the relative calibration, we do not draw any conclusions as to which of the two instruments is better calibrated in an absolute sense. We use the RGS to determine the best fit to the spectrum, because of the high-resolution needed for the precise determination of the outflow properties. 
However as can be seen from Fig. \ref{fig:resrgs}, the RGS spectrum shows broad band fluctuations on the order of 3$-$4 $\%$. As already mentioned earlier, this sets a limit on how good our relative calibration modeling can get. Also the exact \textit{knak} model we have is at this moment strictly valid for sources with the same spectral type as Mrk 509 as there are indications that the relative calibration differences appear to be source and pn mode dependent\footnote{http://xmm2.esac.esa.int/external \\
/xmm\_sw\_cal/calib/cross\_cal/index.php}. 
This indicates that the redistribution of pn is partly responsible for the differences in the RGS and pn flux, since a different spectral slope will have an important effect on the exact redistribution of photons in the pn spectrum. 
This is also what has been reported to the \textit{XMM-Newton} Users Group\footnote{http://xmm.esac.esa.int/external/xmm\_user\_support/ \\ 
         usersgroup/20090506/index.shtml}. 
By applying this method to other sources (including non-AGN sources) and looking at the exact differences between the obtained relative calibration residuals (i.e. the exact values of the \textit{knak} model for example), gives us an opportunity to try to determine the exact origin of the differences between RGS and pn. 

\section{Conclusions} 		\label{conclusions}
We have investigated the cross-calibration differences between \textit{XMM} pn and RGS. By modeling the cross-calibration differences, which are consistent for all three observations and the stacked spectrum, we are able to improve the calibration for pn and RGS to within 4$\%$ of each other. With this improvement we can detect small variations in the outflow column density and ionization state more accurately. 
For Mrk 509 we have used this method in combination with the \textit{RXTE} light curve to constrain the location of the highly ionized component (log $\xi$ $\approx$ 3.2) to within 4 pc of the central source. The other components do not show any significant variability, despite the source varying in flux between the 2005 and 2006 observations.
With this improvement to the relative calibration of the XMM instruments, we can use the advantage of simultaneous pn and RGS observations to detect and model small changes in the outflow of AGN. This will allow us to expand outflow studies to sources which vary more slowly and with a smaller amplitude, thereby increasing the number of Seyfert 1 galaxies that can be studied in this way. This method can also be expanded to other types of sources, as the method itself is source independent. 
 
\begin{acknowledgements}
The authors also want to thank Matteo Guianazzi and Ulrich Briel for advice concerning X-ray loading. We also want to thank Andy Pollock for fruitful discussions regarding the RGS EPIC-pn cross-calibration. This work is based on observations obtained with \textit{XMM-Newton}, an ESA science mission with instruments and contributions directly funded by ESA Member States and the USA (NASA). SRON is supported financially by NWO, the Netherlands Organization for Scientific Research.     
\end{acknowledgements}

\bibliographystyle{aa}
\bibliography{bibfiles}

\begin{thebibliography}{29}
\expandafter\ifx\csname natexlab\endcsname\relax\def\natexlab#1{#1}\fi

\bibitem[{{Behar} {et~al.}(2003){Behar}, {Rasmussen}, {Blustin}, {Sako},
  {Kahn}, {Kaastra}, {Branduardi-Raymont}, \& {Steenbrugge}}]{Behar03}
{Behar}, E., {Rasmussen}, A.~P., {Blustin}, A.~J., {et~al.} 2003, \apj, 598,
  232

\bibitem[{{Bottorff} {et~al.}(2000){Bottorff}, {Korista}, \&
  {Shlosman}}]{Bottorff00}
{Bottorff}, M.~C., {Korista}, K.~T., \& {Shlosman}, I. 2000, \apj, 537, 134

\bibitem[{{Cappi} {et~al.}(2009){Cappi}, {Tombesi}, {Bianchi}, {Dadina},
  {Giustini}, {Malaguti}, {Maraschi}, {Palumbo}, {Petrucci}, {Ponti},
  {Vignali}, \& {Yaqoob}}]{Cappi09}
{Cappi}, M., {Tombesi}, F., {Bianchi}, S., {et~al.} 2009, \aap, 504, 401

\bibitem[{{Cardelli} {et~al.}(1989){Cardelli}, {Clayton}, \&
  {Mathis}}]{Cardelli89}
{Cardelli}, J.~A., {Clayton}, G.~C., \& {Mathis}, J.~S. 1989, \apj, 345, 245

\bibitem[{{Chakravorty} {et~al.}(2009){Chakravorty}, {Kembhavi}, {Elvis}, \&
  {Ferland}}]{Chakravorty09}
{Chakravorty}, S., {Kembhavi}, A.~K., {Elvis}, M., \& {Ferland}, G. 2009,
  \mnras, 393, 83

\bibitem[{{den Herder} {et~al.}(2001){den Herder}, {Brinkman}, {Kahn},
  {Branduardi-Raymont}, {Thomsen}, {Aarts}, {Audard}, {Bixler}, {den Boggende},
  {Cottam}, {Decker}, {Dubbeldam}, {Erd}, {Goulooze}, {G{\"u}del}, {Guttridge},
  {Hailey}, {Janabi}, {Kaastra}, {de Korte}, {van Leeuwen}, {Mauche},
  {McCalden}, {Mewe}, {Naber}, {Paerels}, {Peterson}, {Rasmussen}, {Rees},
  {Sakelliou}, {Sako}, {Spodek}, {Stern}, {Tamura}, {Tandy}, {de Vries},
  {Welch}, \& {Zehnder}}]{denHerder01}
{den Herder}, J.~W., {Brinkman}, A.~C., {Kahn}, S.~M., {et~al.} 2001, \aap,
  365, L7

\bibitem[{{Detmers} {et~al.}(2008){Detmers}, {Kaastra}, {Costantini},
  {McHardy}, \& {Verbunt}}]{Detmers08}
{Detmers}, R.~G., {Kaastra}, J.~S., {Costantini}, E., {McHardy}, I.~M., \&
  {Verbunt}, F. 2008, \aap, 488, 67

\bibitem[{{Di Matteo} {et~al.}(2005){Di Matteo}, {Springel}, \&
  {Hernquist}}]{DiMatteo05}
{Di Matteo}, T., {Springel}, V., \& {Hernquist}, L. 2005, \nat, 433, 604

\bibitem[{{Engstrom} \& {Litzen}(1995)}]{Engstrom95}
{Engstrom}, L. \& {Litzen}, U. 1995, Journal of Physics B Atomic Molecular
  Physics, 28, 2565

\bibitem[{{Ferland} {et~al.}(1998){Ferland}, {Korista}, {Verner}, {Ferguson},
  {Kingdon}, \& {Verner}}]{Ferland98}
{Ferland}, G.~J., {Korista}, K.~T., {Verner}, D.~A., {et~al.} 1998, \pasp, 110,
  761

\bibitem[{{Fisher} {et~al.}(1995){Fisher}, {Huchra}, {Strauss}, {Davis},
  {Yahil}, \& {Schlegel}}]{Fisher95}
{Fisher}, K.~B., {Huchra}, J.~P., {Strauss}, M.~A., {et~al.} 1995, \apjs, 100,
  69

\bibitem[{{Gabel} {et~al.}(2003){Gabel}, {Crenshaw}, {Kraemer}, {Brandt},
  {George}, {Hamann}, {Kaiser}, {Kaspi}, {Kriss}, {Mathur}, {Mushotzky},
  {Nandra}, {Netzer}, {Peterson}, {Shields}, {Turner}, \& {Zheng}}]{Gabel03}
{Gabel}, J.~R., {Crenshaw}, D.~M., {Kraemer}, S.~B., {et~al.} 2003, \apj, 583,
  178

\bibitem[{{Garcia} \& {Mack}(1965)}]{Garcia65}
{Garcia}, J.~D. \& {Mack}, J.~E. 1965, Journal of the Optical Society of
  America (1917-1983), 55, 654

\bibitem[{{Kaastra} {et~al.}(2009){Kaastra}, {Lanz}, {Hubeny}, \&
  {Paerels}}]{Kaastra09}
{Kaastra}, J.~S., {Lanz}, T., {Hubeny}, I., \& {Paerels}, F.~B.~S. 2009, \aap,
  497, 311

\bibitem[{{Kaastra} {et~al.}(1996){Kaastra}, {Mewe}, \&
  {Nieuwenhuijzen}}]{Kaastra96}
{Kaastra}, J.~S., {Mewe}, R., \& {Nieuwenhuijzen}, H. 1996, in UV and X-ray
  Spectroscopy of Astrophysical and Laboratory Plasmas, ed. {K.~Yamashita \&
  T.~Watanabe}, 411--414

\bibitem[{{Kaastra} {et~al.}(2004){Kaastra}, {Raassen}, {Mewe}, {Arav},
  {Behar}, {Costantini}, {Gabel}, {Kriss}, {Proga}, {Sako}, \&
  {Steenbrugge}}]{Kaastra04}
{Kaastra}, J.~S., {Raassen}, A.~J.~J., {Mewe}, R., {et~al.} 2004, \aap, 428, 57

\bibitem[{{Kelly}(1987)}]{Kelly87}
{Kelly}, R.~L. 1987, {Atomic and ionic spectrum lines below 2000 Angstroms.
  Hydrogen through Krypton}, ed. {Kelly, R.~L.}

\bibitem[{{Krolik} \& {Kriss}(1995)}]{Krolik95}
{Krolik}, J.~H. \& {Kriss}, G.~A. 1995, \apj, 447, 512

\bibitem[{{Krongold} {et~al.}(2005){Krongold}, {Nicastro}, {Brickhouse},
  {Elvis}, \& {Mathur}}]{Krongold05}
{Krongold}, Y., {Nicastro}, F., {Brickhouse}, N.~S., {Elvis}, M., \& {Mathur},
  S. 2005, \apj, 622, 842

\bibitem[{{Murphy} {et~al.}(1996){Murphy}, {Lockman}, {Laor}, \&
  {Elvis}}]{Murphy96}
{Murphy}, E.~M., {Lockman}, F.~J., {Laor}, A., \& {Elvis}, M. 1996, \apjs, 105,
  369

\bibitem[{{Netzer} {et~al.}(2002){Netzer}, {Chelouche}, {George}, {Turner},
  {Crenshaw}, {Kraemer}, \& {Nandra}}]{Netzer02}
{Netzer}, H., {Chelouche}, D., {George}, I.~M., {et~al.} 2002, \apj, 571, 256

\bibitem[{{Netzer} {et~al.}(2003){Netzer}, {Kaspi}, {Behar}, {Brandt},
  {Chelouche}, {George}, {Crenshaw}, {Gabel}, {Hamann}, {Kraemer}, {Kriss},
  {Nandra}, {Peterson}, {Shields}, \& {Turner}}]{Netzer03}
{Netzer}, H., {Kaspi}, S., {Behar}, E., {et~al.} 2003, \apj, 599, 933

\bibitem[{{Page} {et~al.}(2003){Page}, {Davis}, \& {Salvi}}]{Page03}
{Page}, M.~J., {Davis}, S.~W., \& {Salvi}, N.~J. 2003, \mnras, 343, 1241

\bibitem[{{Pounds} {et~al.}(2001){Pounds}, {Reeves}, {O'Brien}, {Page},
  {Turner}, \& {Nayakshin}}]{Pounds01}
{Pounds}, K., {Reeves}, J., {O'Brien}, P., {et~al.} 2001, \apj, 559, 181

\bibitem[{{Predehl} \& {Schmitt}(1995)}]{Predehl95}
{Predehl}, P. \& {Schmitt}, J.~H.~M.~M. 1995, \aap, 293, 889

\bibitem[{{Smith} {et~al.}(2007){Smith}, {Page}, \&
  {Branduardi-Raymont}}]{Smith07}
{Smith}, R.~A.~N., {Page}, M.~J., \& {Branduardi-Raymont}, G. 2007, \aap, 461,
  135

\bibitem[{{Steenbrugge} {et~al.}(2009){Steenbrugge}, {Fenov{\v c}{\'{\i}}k},
  {Kaastra}, {Costantini}, \& {Verbunt}}]{Steenbrugge09}
{Steenbrugge}, K.~C., {Fenov{\v c}{\'{\i}}k}, M., {Kaastra}, J.~S.,
  {Costantini}, E., \& {Verbunt}, F. 2009, \aap, 496, 107

\bibitem[{{Str{\"u}der} {et~al.}(2001){Str{\"u}der}, {Briel}, {Dennerl},
  {Hartmann}, {Kendziorra}, {Meidinger}, {Pfeffermann}, {Reppin}, {Aschenbach},
  {Bornemann}, {Br{\"a}uninger}, {Burkert}, {Elender}, {Freyberg}, {Haberl},
  {Hartner}, {Heuschmann}, {Hippmann}, {Kastelic}, {Kemmer}, {Kettenring},
  {Kink}, {Krause}, {M{\"u}ller}, {Oppitz}, {Pietsch}, {Popp}, {Predehl},
  {Read}, {Stephan}, {St{\"o}tter}, {Tr{\"u}mper}, {Holl}, {Kemmer}, {Soltau},
  {St{\"o}tter}, {Weber}, {Weichert}, {von Zanthier}, {Carathanassis}, {Lutz},
  {Richter}, {Solc}, {B{\"o}ttcher}, {Kuster}, {Staubert}, {Abbey}, {Holland},
  {Turner}, {Balasini}, {Bignami}, {La Palombara}, {Villa}, {Buttler},
  {Gianini}, {Lain{\'e}}, {Lumb}, \& {Dhez}}]{Struder01}
{Str{\"u}der}, L., {Briel}, U., {Dennerl}, K., {et~al.} 2001, \aap, 365, L18

\bibitem[{{Yaqoob} {et~al.}(2003){Yaqoob}, {McKernan}, {Kraemer}, {Crenshaw},
  {Gabel}, {George}, \& {Turner}}]{Yaqoob03}
{Yaqoob}, T., {McKernan}, B., {Kraemer}, S.~B., {et~al.} 2003, \apj, 582, 105

\end{thebibliography}
\newpage

\appendix

\section{Additional figures}

\begin{figure*}[tbp]
\centering
\hbox{
   \includegraphics[angle= -90,width=8cm]{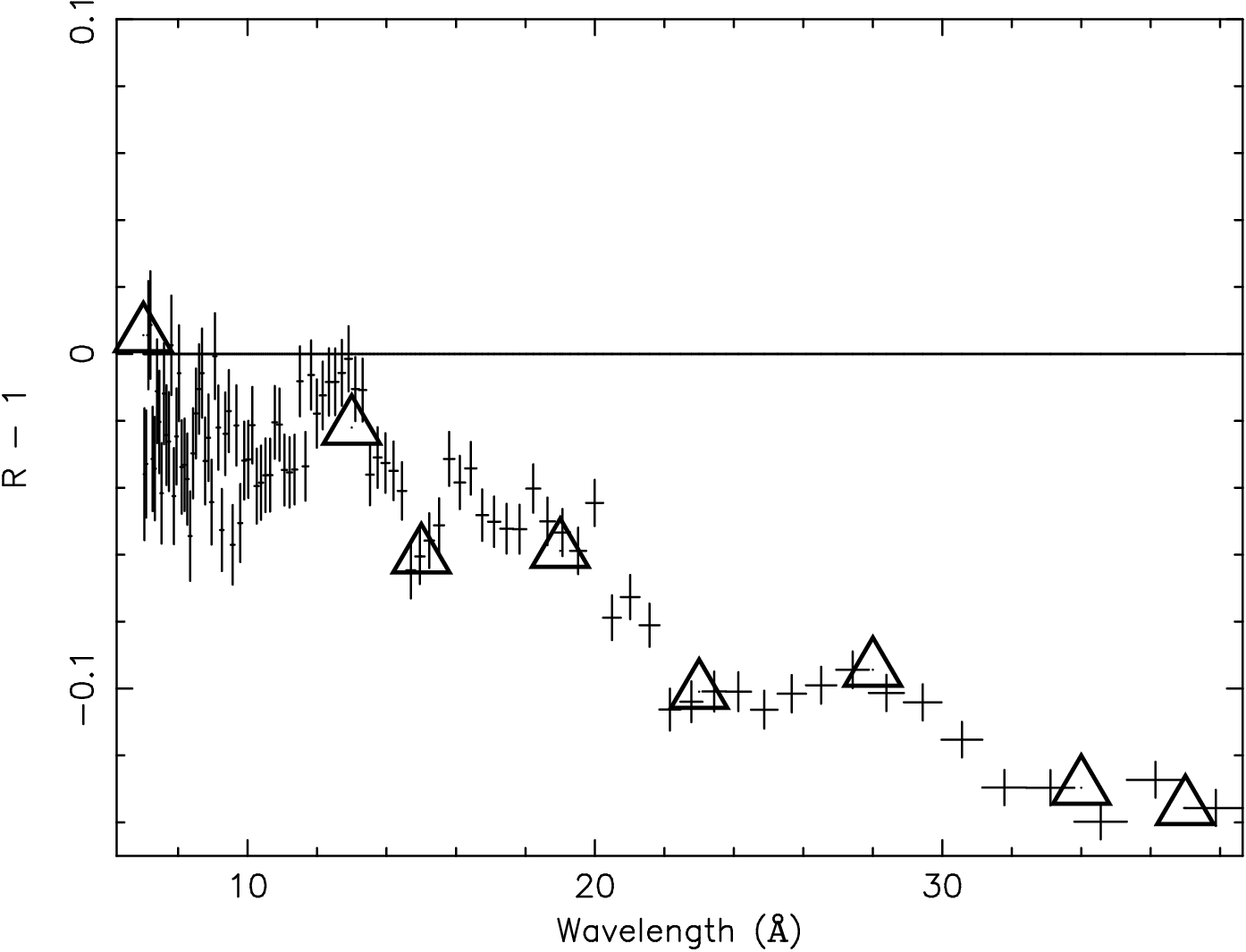}
   \includegraphics[angle= -90,width=8cm]{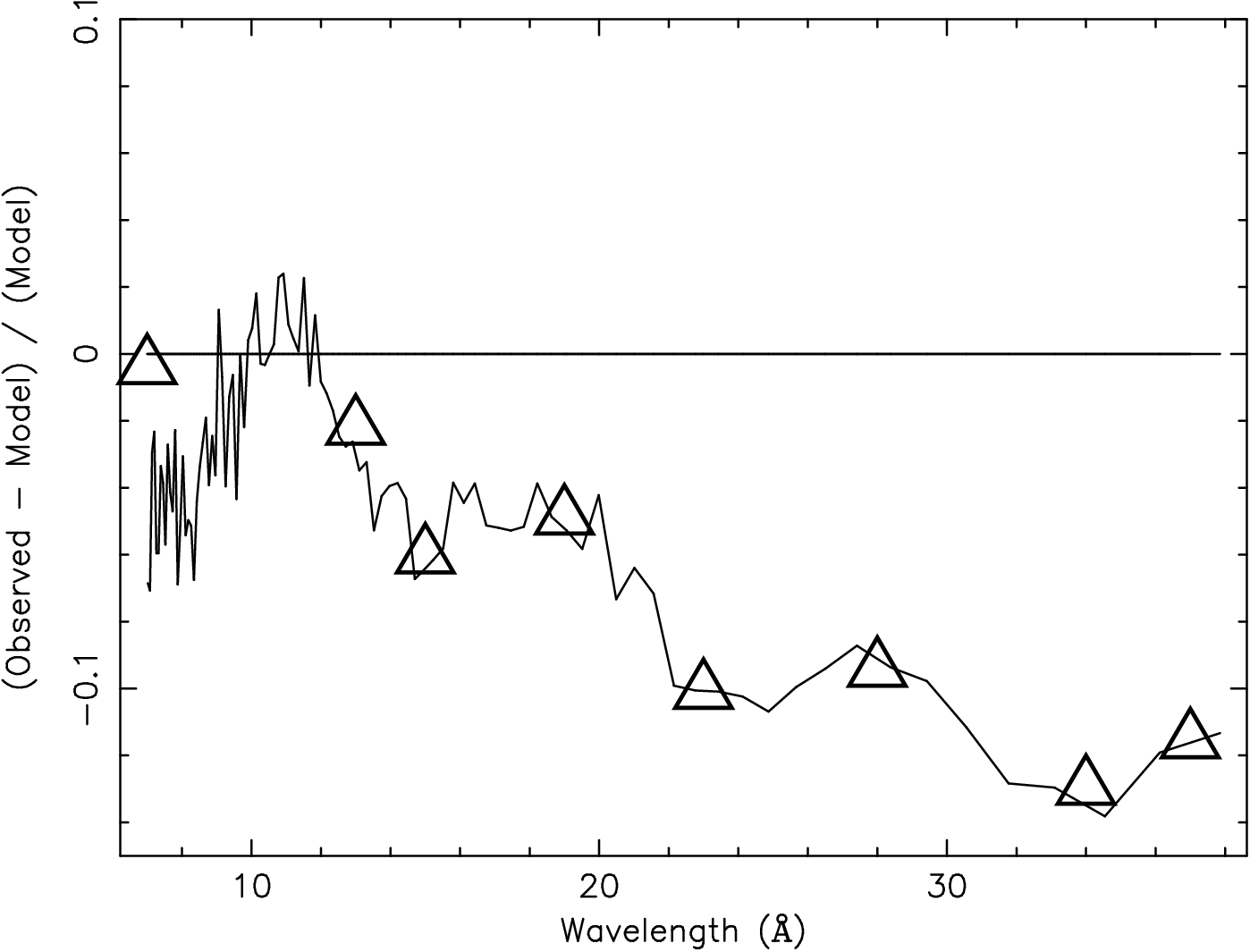}
   }
\hbox{   
   \includegraphics[angle= -90,width=8cm]{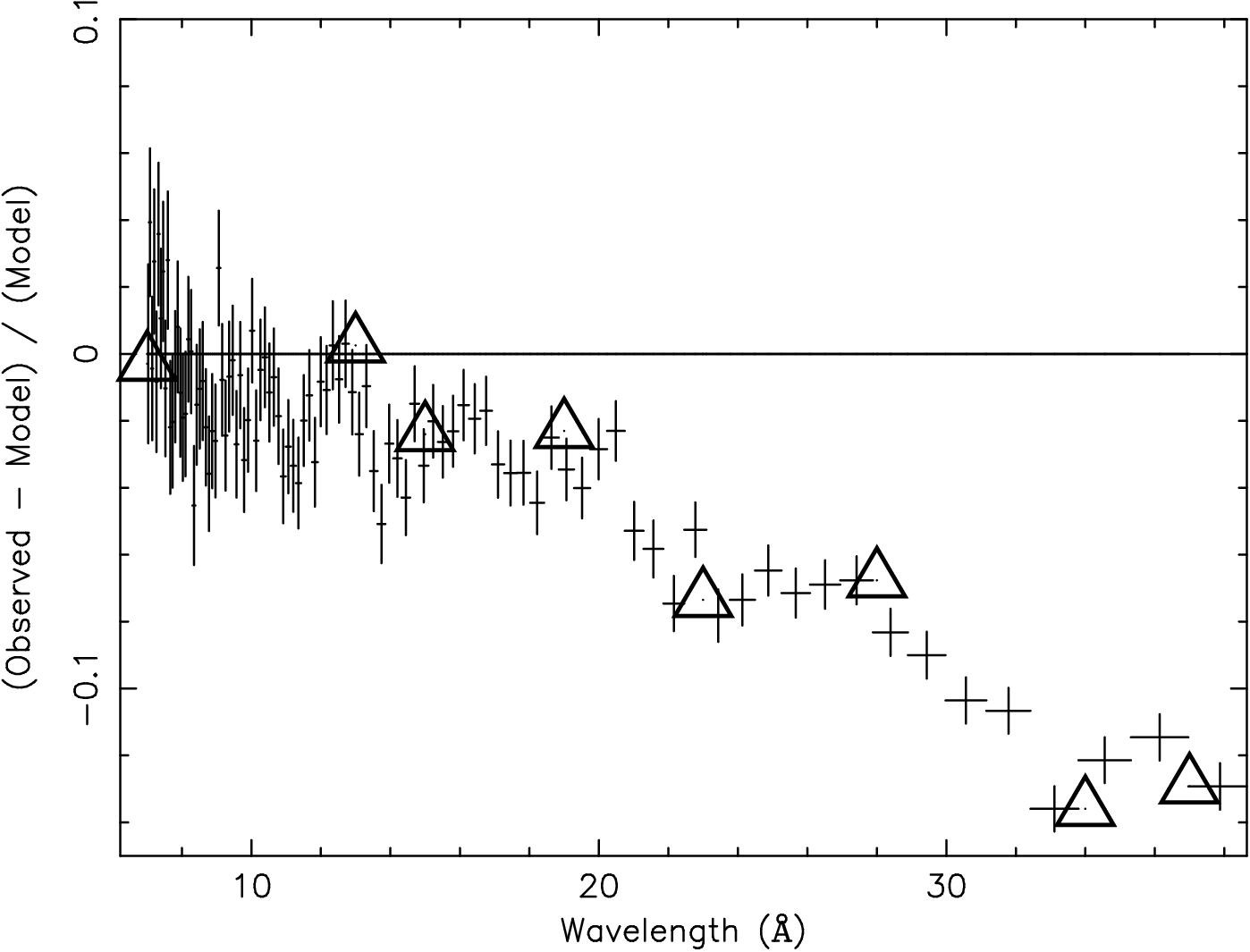}
   \includegraphics[angle= -90,width=8cm]{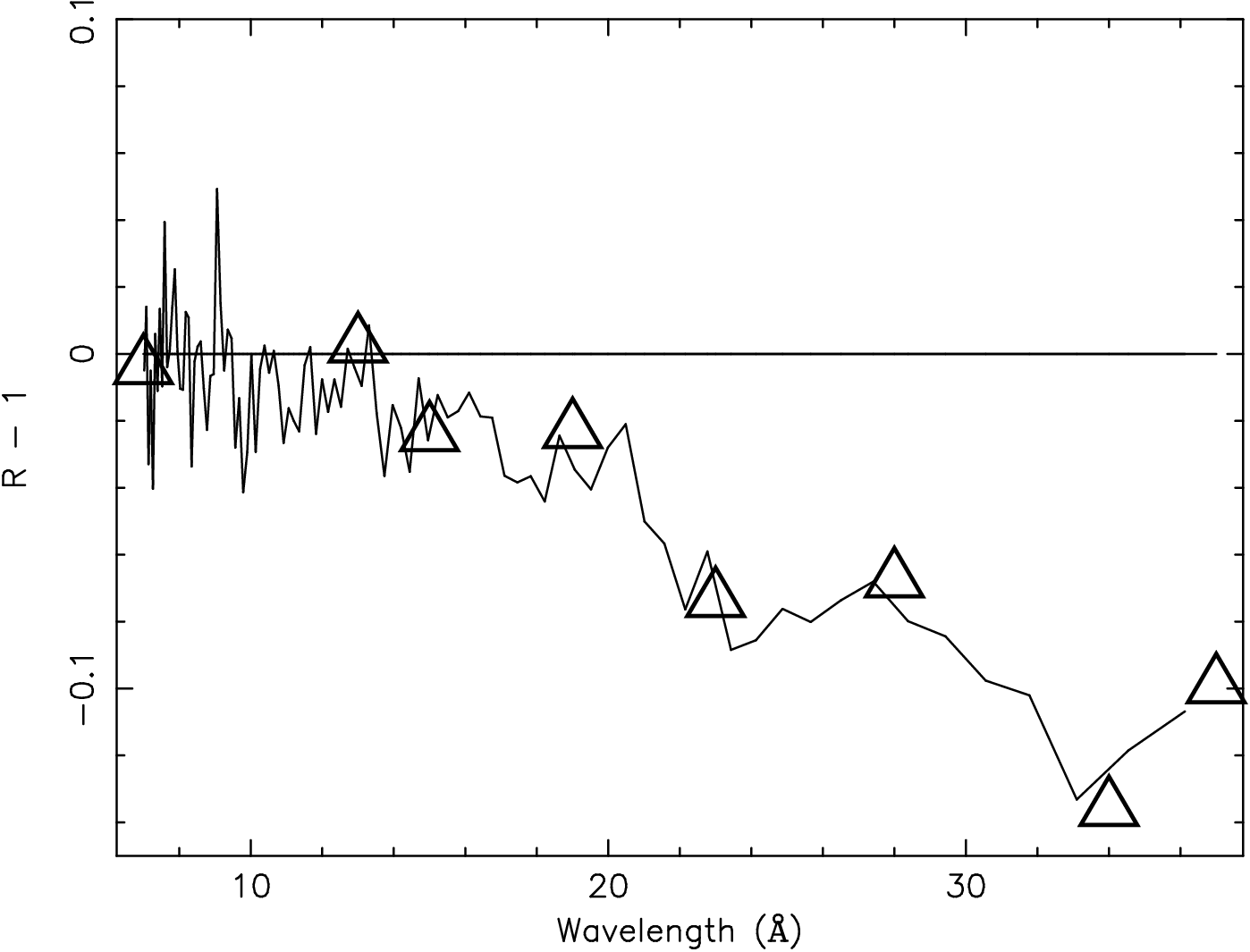}
}
\hbox{   
   \includegraphics[angle= -90,width=8cm]{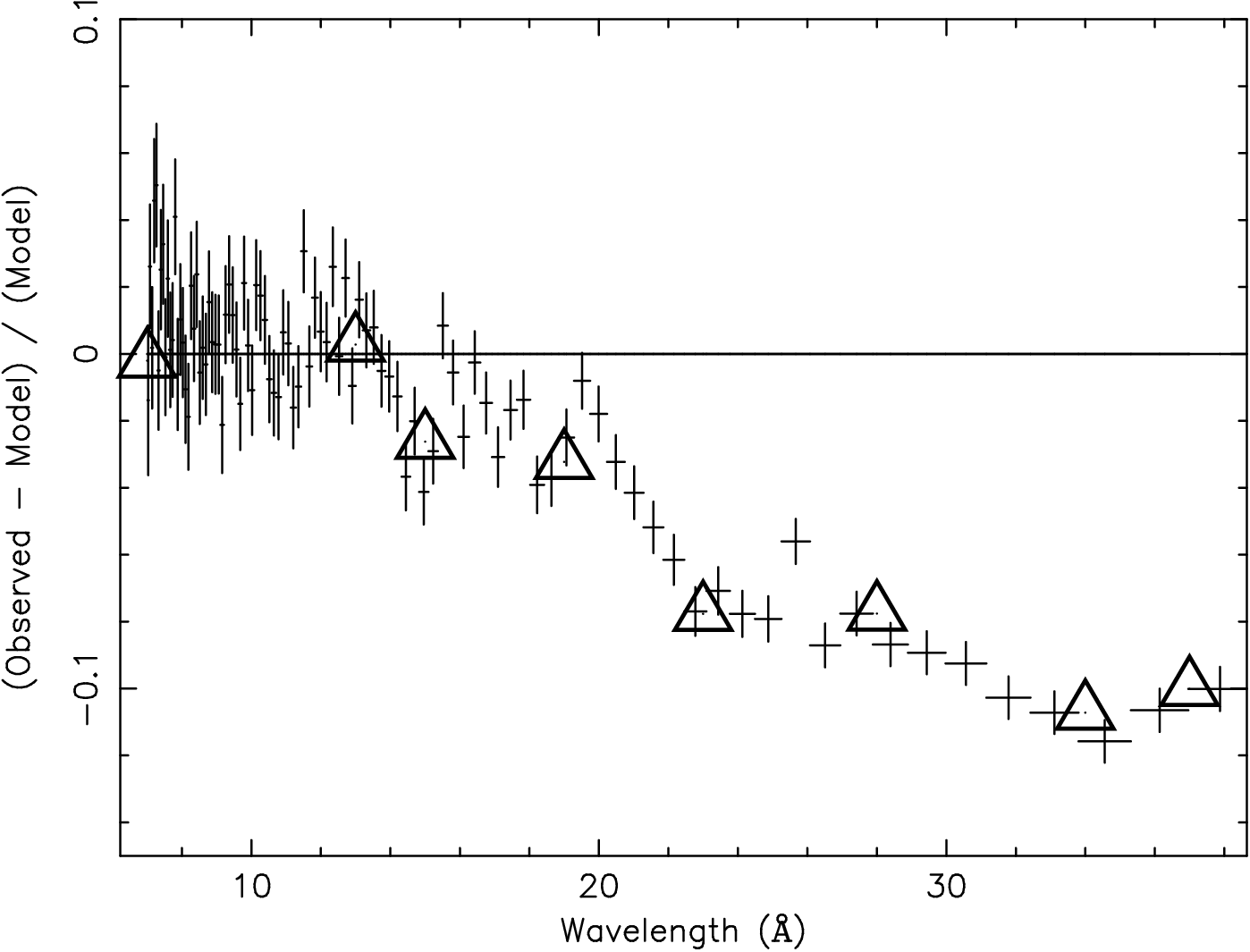}
   \includegraphics[angle= -90,width=8cm]{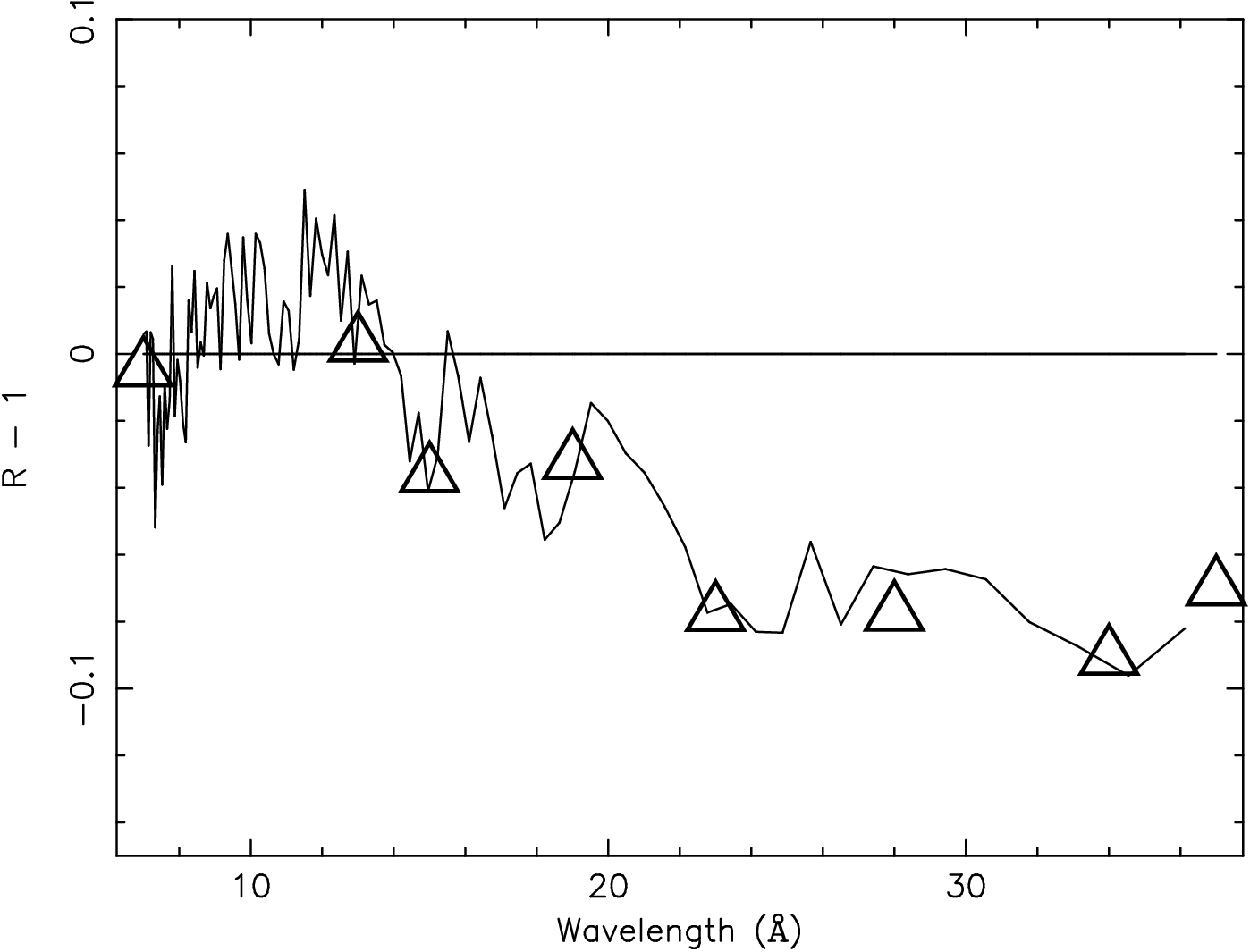}
}
\hbox{   
   \includegraphics[angle= -90,width=8cm]{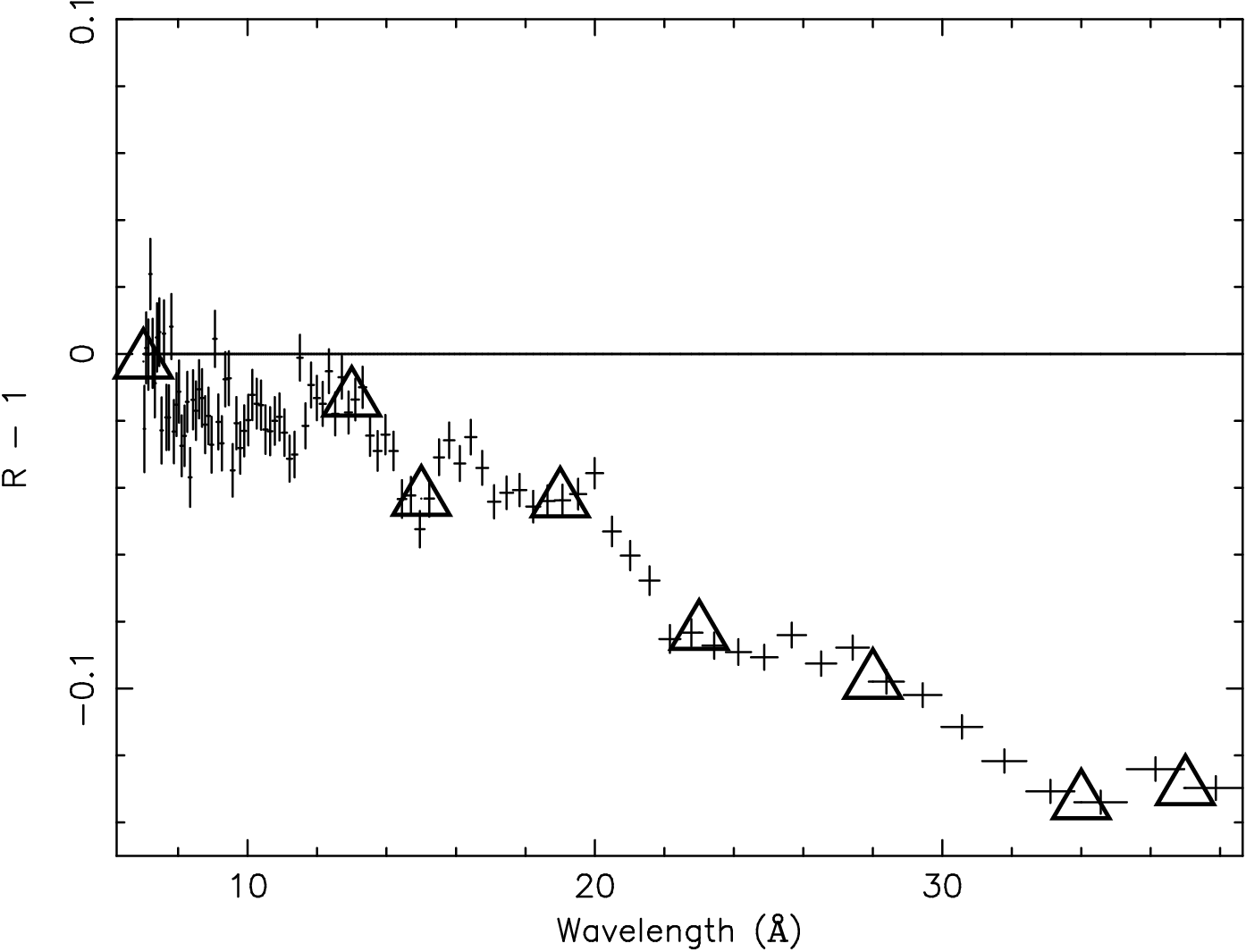}
   \includegraphics[angle= -90,width=8cm]{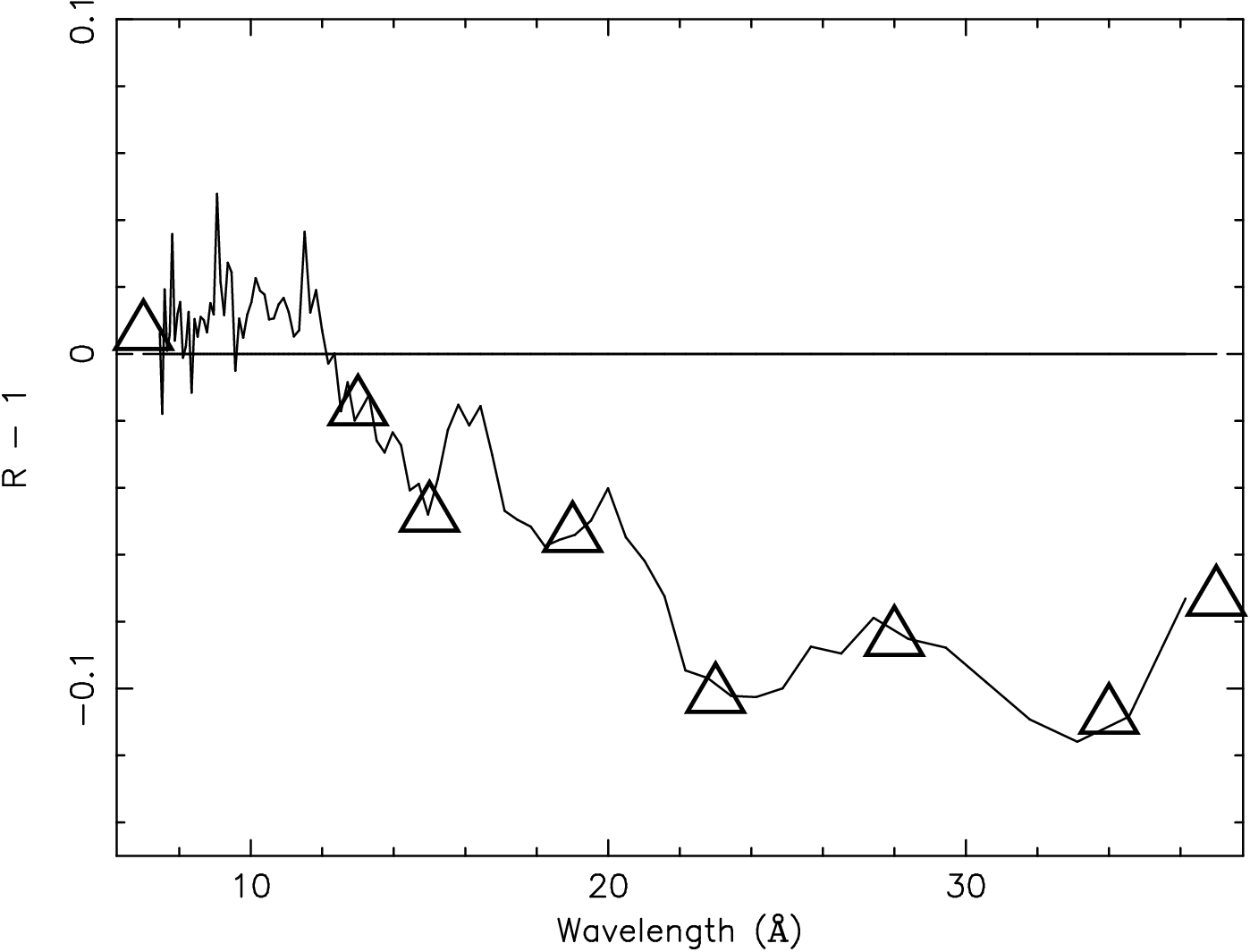}
}   
   \caption{\label{fig:resepic}
        The ratios $R(PI)$ = $\frac{C_{O}(PI)}{C_{M}(PI)}$ for all observations. The left-hand column shows the ratios for method 1, while the right-hand column shows the ratios for method 2. The locations of the \textit{knak} hinge points are indicated by the triangles (The y-axis values for each point are not accurate, as the values shown in Table \ref{tab:knak} are averaged between hinge points, the triangles shown here are just to guide the eye). From top to bottom: Observation 1, 2, 3 and the stacked spectrum. }
\end{figure*}

\begin{figure}[tbp]
   \includegraphics[angle= -90,width=8cm]{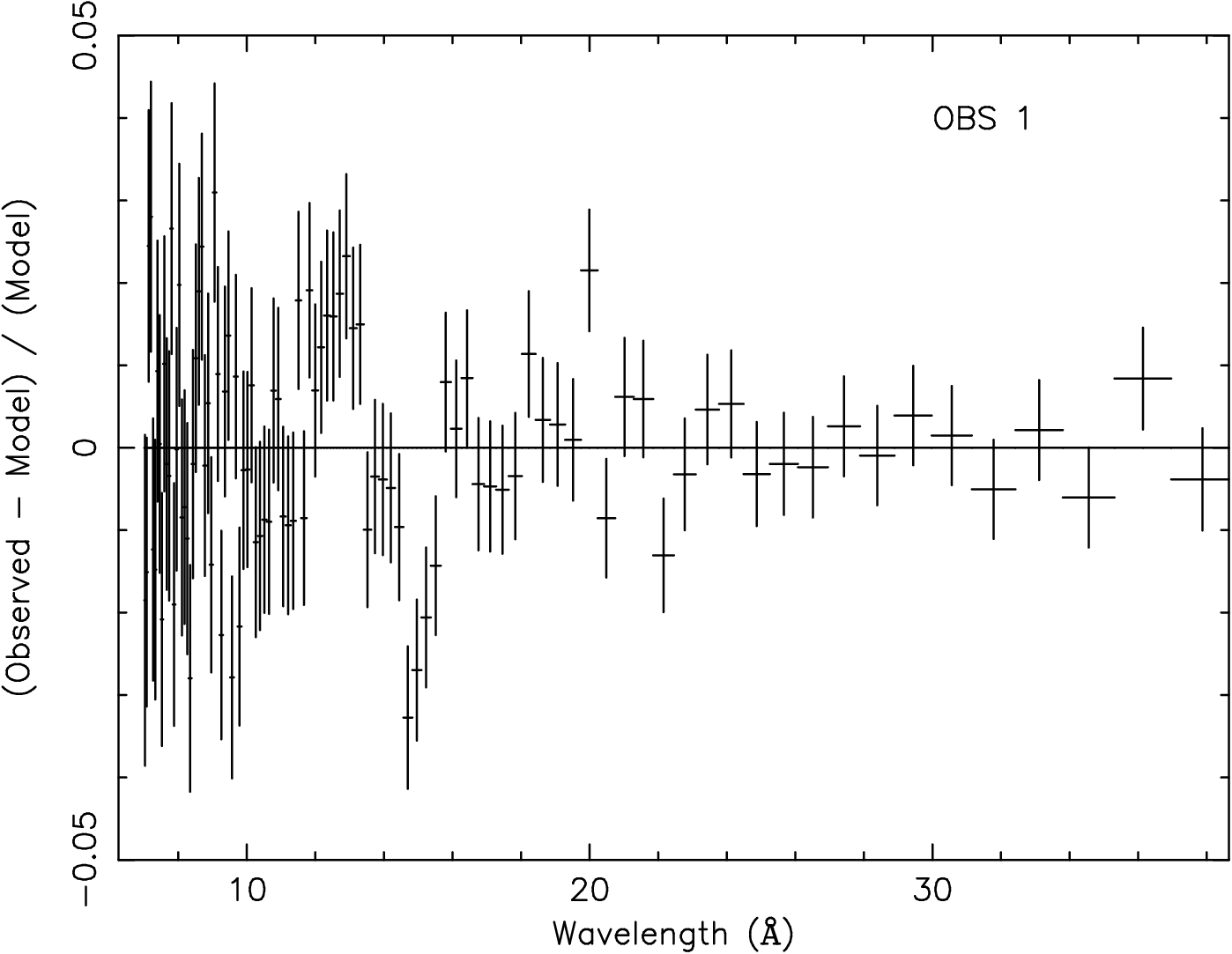}
   \includegraphics[angle= -90,width=8cm]{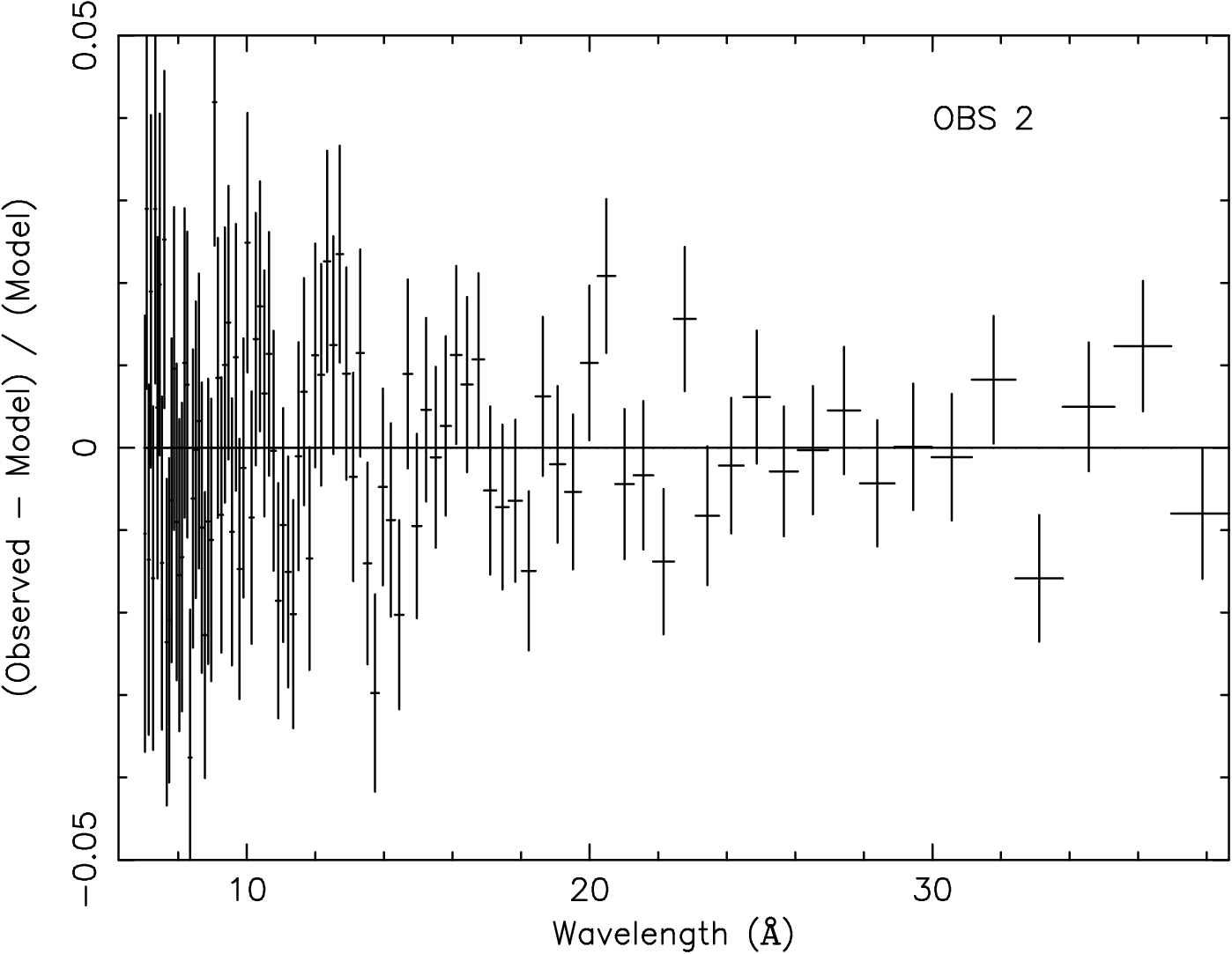}
   \includegraphics[angle= -90,width=8cm]{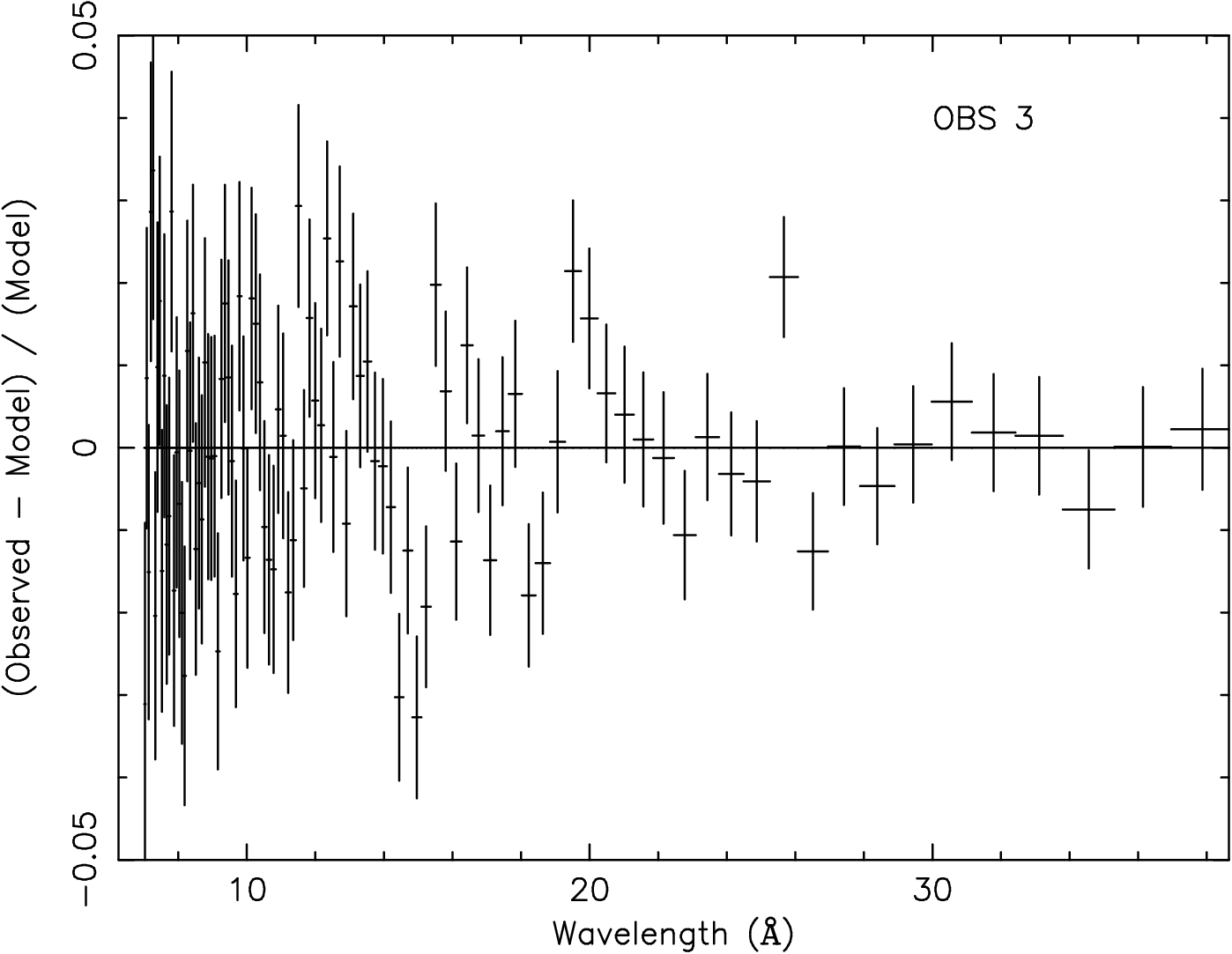}
   \includegraphics[angle= -90,width=8cm]{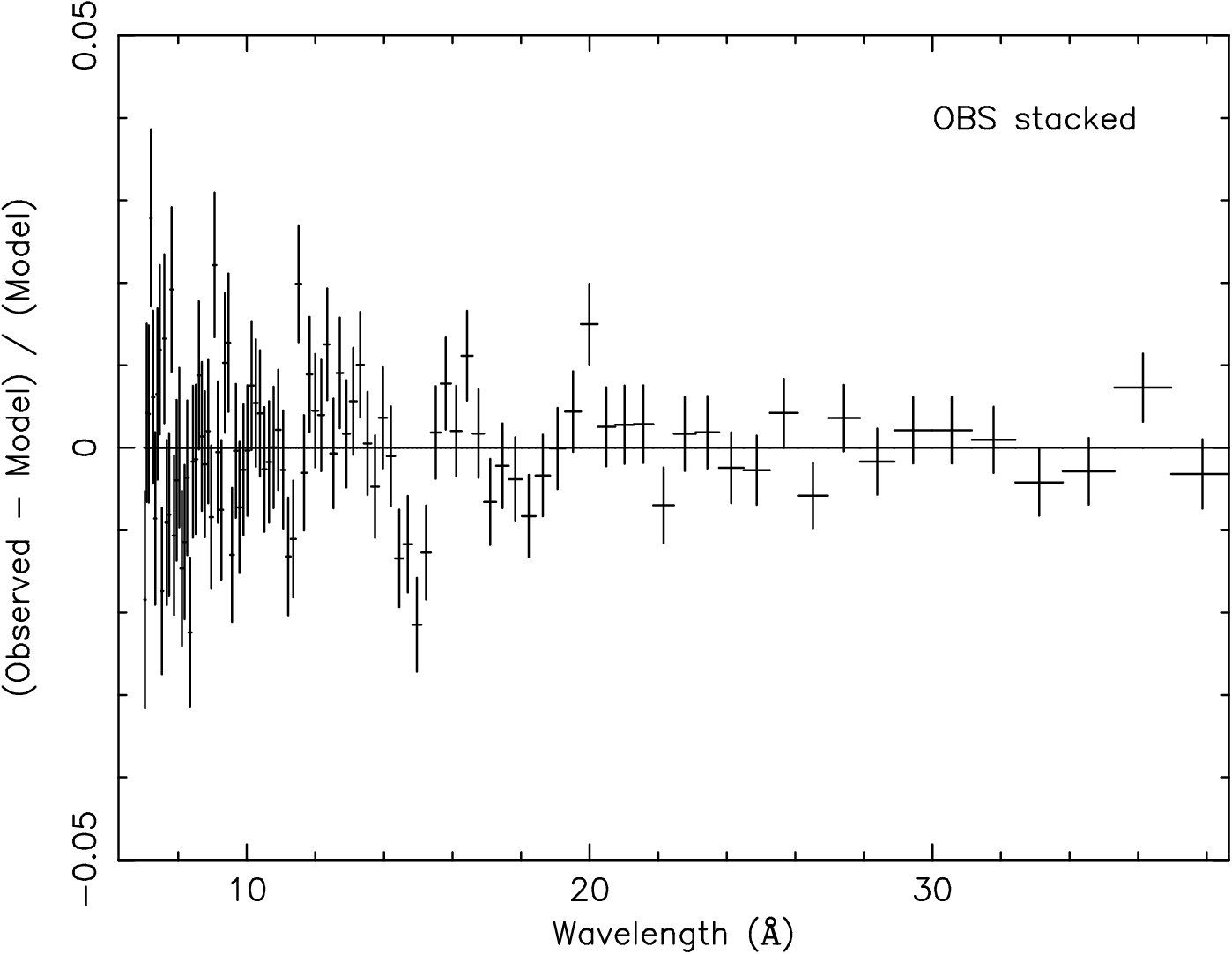}
   \caption{\label{fig:knakres}
        The best fit residuals for the pn spectra after applying the \textit{knak} model. From top to bottom: Observation 1, 2, 3 and the stacked spectrum.
             }
\end{figure}

\end{document}